\documentclass[aps,prd,floatfix,nofootinbib,superscriptaddress,twocolumn,showkeys]{revtex4-2}
\usepackage{amsmath}
\usepackage{graphicx} 
\usepackage{color}
\usepackage{physics}
\usepackage{verbatim}
\usepackage{tikz}
\usepackage{amssymb}
\usepackage{listings}
\usepackage{amssymb}
\allowdisplaybreaks
\usepackage{float}
\usepackage{makeidx}
\usepackage{amsfonts}
\usepackage[usenames,dvipsnames]{pstricks}
\usepackage{subfigure}
\usepackage{epsfig}
\usepackage{pst-grad} 
\usepackage{mathtools}
\usepackage[colorlinks,hyperindex]{hyperref}
\usepackage{array}
\usepackage{color}
\hypersetup
{	colorlinks,
	citecolor=black,
	linkcolor=black,
	urlcolor=black,
}
\usepackage{allpurpose}
\usepackage{flushend}
\usepackage[utf8]{inputenc}
\newcommand\nn{{\nonumber}}

\def\XXint#1#2#3{{\setbox0=\hbox{$#1{#2#3}{\int}$}	\vcenter{\hbox{$#2#3$}}\kern-.5\wd0}}
\usepackage{listings}
\usepackage{ulem}

\usepackage{color}

\newcommand{\delred}[1]{{\color{red}{\ifmmode\text{\sout{\ensuremath{#1}}}\else\sout{#1}\fi}}}

\begin{document}

\title{Off-equatorial deflections and gravitational lensing. II. In general stationary and axisymmetric spacetimes}

\author{Xinguang Ying}
\affiliation{School of Physics and Technology, Wuhan University, Wuhan, 430072, China}

\author{Junji Jia}
\email[Corresponding author:~]{junjijia@whu.edu.cn}
\affiliation{Department of Astronomy \& MOE Key Laboratory of Artificial Micro- and Nano-structures, School of Physics and Technology, Wuhan University, Wuhan, 430072, China}

\date{\today}

\begin{abstract}
In this work, we develop a general perturbative procedure to find the off-equatorial plane deflections in the weak deflection limit in general stationary and axisymmetric spacetimes, allowing the existence of the generalized Carter constant. Deflections of both null and timelike rays, with the finite distance effect of the source and detector taken into account, are obtained as dual series of $M/r_0$ and $r_0/r_{s,d}$. These deflections allow a set of exact gravitational lensing equations from which the images' apparent angular positions are solved. The method and general results are then applied to the Kerr-Newmann, Kerr-Sen, and rotating Simpson-Visser spacetimes to study the effect of the spin and characteristic (effective) charge of the spacetimes and the source altitude on the deflection angles and image apparent angles. It is found that, in general, both the spacetime spin and charge only affect the deflections from the second non-trivial order, while the source altitude influences the deflection from the leading order. Because of this, it is found that, in gravitational lensing in realistic situations, it is hard to measure the effects of the spacetime spin and charge from the images' apparent locations. We also presented the off-equatorial deflections in the rotating Bardeen, Hayward, Ghosh, and Tinchev black hole spacetimes.  
\end{abstract}


\maketitle

\section{Introduction \label{sec:Introduction}}

Deflection of light rays in gravity has been extensively studied from the very early stage of General Relativity \cite{Dyson:1920cwa,Einstein:1936llh}. Nowadays, gravitational lensing (GL) has developed into a powerful tool in astronomy, ranging from measuring the mass of galaxies or their clusters \cite{Bartelmann:1999yn}, studying distributions of dark matter \cite{Metcalf:2001ap}, properties of supernovas \cite{Sharon:2014ija} and even testing alternative gravitational theories \cite{Keeton:2005jd,Joyce:2016vqv}.

The simplest scenario of signal deflection and GL is that of light rays in static and spherically symmetric (SSS) spacetimes, or in the equatorial plane of stationary and axisymmetric (SAS) spacetimes in the weak deflection limit (WDL). With the fast developments of astroparticle physics \cite{AMS:2021nhj,Bionta:1987qt}, gravitational wave detection \cite{LIGOScientific:2016aoc}, and black hole (BH) imaging \cite{EventHorizonTelescope:2019dse,EventHorizonTelescope:2022wkp}, enormous effort has been devoted to the extension of the deflection and GL of timelike signals \cite{Jia:2017nen,Jia:2017jff,Glicenstein:2017lrm,Crisnejo:2019ril,Li:2019vhp,Li:2020dln}, with finite source and detector distance \cite{Ishihara:2016vdc}, and in the strong deflection limit \cite{Bozza:2002zj,Zhou:2022dze,Jia:2020qzt}. Different analytical methods were also developed, including the perturbative methods \cite{Jia:2020xbc,Huang:2020trl,Jia:2020qzt,Duan:2023gvm} and the more recent Gauss-Bonnet theorem-based methods \cite{Gibbons:2008rj,Crisnejo:2019ril,Li:2019vhp,Li:2020dln}. 

However, extension to the non-equatorial deflection and GL in SAS spacetime is still rare (in SSS spacetime essentially there is no non-equatorial motion), except in Kerr \cite{Fujita:2009bp, Bray:1985ew,Sereno:2006ss,Kraniotis:2010gx, Gralla:2019drh,Yang:2013yoa,Jiang:2023yvo} and Kerr-Newmann (KN) \cite{Hackmann:2013pva,Yang:2013wya} spacetime. Due to the complexity of the motion equations for the off-equatorial trajectories, only a few works have studied the deflection and GL in quasi-equatorial motion of null rays \cite{Gyulchev:2006zg,Molla:2021sgw} or only numerically in the general non-equatorial case \cite{Wang:2017hjl,Islam:2021ful} in other spacetimes, let alone the deflection of both null and timelike rays. In Ref. \cite{Jiang:2023yvo}, the general non-equatorial deflection and GL in Kerr spacetime were studied perturbatively for the first time for both null and timelike rays in the WDL, with finite distance effect taken into account. One of the motivations of the current work is to study the condition under the form of SAS spacetime for the perturbative method to be feasible for non-equatorial deflection and GL. We will show that for many SAS spacetimes satisfying a separation condition that allows a generalized Carter constant (GCC), the perturbative method is always valid for both null and timelike rays with the finite distance effect automatically taken into account. The second motivation is to reveal the effects of various spacetime parameters on such deflections and GL. Spacetimes will be considered include the KN, Kerr-Sen (KS) \cite{Sen:1992ua}, rotating Simpson–Visser (RSV) \cite{Mazza:2021rgq} and a few other ones \cite{Bambi:2013ufa,Ghosh:2014pba,Tinchev:2015apf,Konoplya:2016pmh}.

The paper is organized as follows. 
In Sec. \ref{sec:Equations of Motion and Deflection}, we first introduce the basic setup of the problem, establish the equations of motion, and study a condition for the perturbative method to work. In Sec. \ref{sec:The Perturbative Method}, we explore a perturbative method to solve the deflection angle in both $\theta$ and $\phi$ direction. The off-equatorial GL equations are then solved in Sec. \ref{sec:Gravitational Lensing Effects} to obtain the apparent angles of the lensed images. The method and deflection and GL results are then applied to the KN, KS, and RSV spacetimes in Sec. \ref{sec:Applications}, and the effects of their characteristic parameters are studied. We conclude the paper with a summary and discussion in Sec. \ref{sec:sumdis}. Throughout the paper, we use the natural unit $G=c=1$ and the spacetime signature ($-,+,+,+$).

\section{General framework}
In this section, we will derive the deflection angles in the $\theta$ and $\phi$ directions for SAS spacetimes. We will show that this is always possible when the metric functions satisfy certain conditions, such that a proper separation of variables in the equations of motion, or equivalently the existence of a GCC, can be accomplished. 

\subsection{Preliminaries \label{sec:Equations of Motion and Deflection}}
We start from the most general SAS spacetime, whose metric can always be expressed in the following form 
\begin{equation}
    {\mathrm{d}s}^2=-A{\mathrm{d}t}^2+B{\mathrm{d}t\mathrm{d}\phi}+C{\mathrm{d}\phi}^2+D{\mathrm{d}r}^2+F{\mathrm{d}\theta}^2,\label{eq:general metric}
\end{equation}
where $t,\,r,\,\theta,\,\phi$ are the Boyer-Lindquist coordinates and $A,\,B,\,C,\,D,\,F$ are functions of $r$ and $\theta$ only. 
This metric allows two commutative Killing vectors $$\xi^\mu=\left(\frac{\partial}{\partial t}\right)^\mu,\,\psi^\mu=\left(\frac{\partial}{\partial \phi}\right)^\mu,\,\left[\xi,\psi\right]^\mu=0,$$
where the spacelike $\psi^\mu$ corresponding to the rotation symmetry and timelike $\xi^\mu$ to the time translation. These Killing vectors correspond to two conserved quantities of the motion
\begin{align}
&E=A\dot{t}-\frac{1}{2}B\dot{\phi},\label{eq:energy}\\
&L=\frac{1}{2}B\dot{t}+C\dot{\phi}.\label{eq:angle momentum}
\end{align}
Here the dot stands for the derivative to the proper time or affine parameter $\lambda$ of the motion and $E$ and $L$ can be interpreted as the energy and angular momentum of the particle (per unit mass) respectively. In asymptotically flat spacetimes, $E$ can also be related to the asymptotic velocity $v$ of the particle through 
\begin{align}
    E=\frac{1}{\sqrt{1-v^2}}. \label{eq:evrel} 
\end{align}
The asymptotic velocity $v$ here stands for the magnitude of the spatial component of the four-velocity of the test particle. From these equations, one can obtain two first derivatives
\begin{align}
    \dot{t}&=\frac{2BL+4EC}{B^2+4AC},
 \label{eq:firstintegralt}\\
    \dot{\phi}&=\frac{4AL-2BE}{B^2+4AC}.
    \label{eq:firstintegralphi}
\end{align}

Now for the equations of motion of the $r$ and $\theta$ coordinates, we can simply write out their geodesic equations. However, they are second-order equations that are very complicated to simplify. In this work, we will limit our choices of the SAS spacetimes, i.e., putting conditions on the metric functions, such that the motions allow a third conserved constant, i.e. the GCC \cite{Bezdekova:2022gib}. 
We note that unlike the Kerr spacetime case, which not only contains a Carter constant \cite{Carter:1968rr} but even second-order Killing tensors \cite{Brink:2009mq}, the existence of GCC is not guaranteed in all SAS spacetimes. For those SAS spacetime without GCC, many of them are non-integrable systems and even the geodesics are chaotic, such as Johannsen-Psaltis spacetime \cite{Zelenka:2017aqn}, Zipoy-Voorhees spacetime \cite{Lukes-Gerakopoulos:2012qpc} and so on. We will not study these spacetimes in this work. 

To see what the existence of such GCC requires, and to obtain (simpler) equations of motion for $r$ and $\theta$, we will use the Hamiltonian-Jacobian approach. Our starting point is the action of the free particle for a separable solution, which reads
\begin{equation}
    S=-\frac{1}{2}\kappa\lambda-Et+L\phi+S^{(r)}(r)+S^{(\theta)}(\theta) ,\label{eq:the action of the free particle}
\end{equation}
where $\kappa=0,\,-1$ for null and timelike particles respectively. Here and hereafter, any function with an $(r)$ or $(\theta)$ superscript is a function of $r$ or $\theta$ only. 
The Hamilton-Jacobi equation is given by
\begin{equation}
\frac{1}{2}g^{\mu\nu}\frac{\partial S}{\partial x^\mu}\frac{\partial S}{\partial x^\nu}-\frac{1}{2}\kappa=0. \label{eq:Hamilton-Jacobi equation}
\end{equation}
Substituting Eq. \eqref{eq:the action of the free particle}, this becomes
\begin{align}
    &\frac{1}{D}\left(\frac{\mathrm{d}S^{(r)}}{\mathrm{d}r}\right)^2+\frac{1}{F}\left(\frac{\mathrm{d}S^{(\theta)}}{\mathrm{d}\theta}\right)^2-\kappa\nn\\&+\frac{4AL^2-4BEL-4CE^2}{B^2+4AC}=0\label{eq:introduce G}.
\end{align}
We then seek metrics that allow this equation, after multiplied by some proper total factor function $G(r,\theta)$, to be separable into $r$ and $\theta$ dependent parts.
This can be accomplished if the metric functions $A,\,B,\,C,\,D,\,F$, and factor $G$ can cast the left-hand sides of the following equations into their right-hand sides \cite{Bezdekova:2022gib}
\begin{subequations}
\label{eq:sepcond}
\begin{align}
&\frac{G(r,\theta)}{D(r,\theta)}=\mathcal{D}(r),~\frac{G(r,\theta)}{F(r,\theta)}=\mathcal{F}(\theta),\label{eq:decompositionF}\\
&\frac{X(r,\theta)G(r,\theta)}{B^2+4AC}\equiv X^{(r)}(r)+X^{(\theta)}(\theta)\label{eq:decompositionX},\\
&G(r,\theta)= G^{(r)}(r)+G^{(\theta)}(\theta),\label{eq:decompositionG}
\end{align}
\end{subequations}
where $X\in\{A,\,B,\,C\}$. 
Note here that the condition \eqref{eq:decompositionG} is for the separability of the $\kappa=-1$ case and unnecessary for null signals. 
Condition \eqref{eq:decompositionF} implies 
\begin{align}
    \frac{F(r,\theta)}{D(r,\theta)}=\frac{\mathcal{D}(r)}{\mathcal{F}(\theta)}. \label{eq:dfcond}
\end{align}
In practice, the functions on the right-hand sides of Eq. \eqref{eq:sepcond} as well as $G(r,\theta)$ can be read off from the left-hand sides and Eq. \eqref{eq:dfcond}. One might also notice that there is a freedom of a multiplicative constant in functions $\mathcal{D}(r)$ and $\mathcal{F}(\theta)$, and additive constant freedom in each pair of functions $X^{(r)}$ and $X^{(\theta)}$. Indeed one can show that these freedoms will be canceled out in the final equations of motion \eqref{eq:motion of r} and \eqref{eq:motion of theta} and therefore not affect the physics. Let us also point out that many SAS spacetimes, including the Kerr spacetime, satisfy these conditions \eqref{eq:sepcond}. 

 A few comments about the variable separation condition \eqref{eq:sepcond} might be useful for their clear understanding here. 
What will be shown in this work is that the spacetimes satisfying the condition \eqref{eq:sepcond} can always be treated using our method, while those spacetimes not satisfying \eqref{eq:sepcond} are not treatable by the method developed in this work.
In this sense, condition \eqref{eq:sepcond} is both a sufficient and necessary condition for the applicability of our method. 
However, Eq. \eqref{eq:sepcond} is only a sufficient condition for the separability of the equations of motion and we are not able to prove that it is also a necessary condition, although we can not provide any counter-example either. That is to say, it is unclear to us whether there exist spacetimes (unknown to us) that do not satisfy condition \eqref{eq:sepcond} but still allow the separation of its variables. 

Using conditions \eqref{eq:sepcond} in Eq. \eqref{eq:introduce G} and separating the $r$ and $\theta$ dependent parts, we obtain
\begin{align}
    &4L^2A^{(r)}-\kappa G^{(r)}-4ELB^{(r)}-4E^2C^{(r)}+\mathcal{D}(r)\left(\frac{\mathrm{d}S^{(r)}}{\mathrm{d}r}\right)^2\nn\\=&\kappa G^{(\theta)}-4L^2A^{(\theta)}+4ELB^{(\theta)}+4E^2C^{(\theta)}-\mathcal{F}(\theta)\left(\frac{\mathrm{d}S^{(\theta)}}{\mathrm{d}\theta}\right)^2\nn\\\equiv& K\label{eq:definition of GCC},
\end{align}
where the assigned constant $K$ is the GCC we are looking for. Note that this GCC also allows some constants because we can always add or multiply some constant to both sides of the first equal sign in Eq. \eqref{eq:definition of GCC}. However, these additive or multiplicative constants will not affect the dynamics and therefore can be chosen freely. Eq. \eqref{eq:definition of GCC} then can be split into two equations
\begin{align}
    \left(\frac{\mathrm{d}S^{(r)}}{\mathrm{d}r}\right)^2&=\frac{\kappa G^{(r)}-4L^2A^{(r)}+4E^2C^{(r)}+4ELB^{(r)}+K}{\mathcal{D}(r)}\nn\\&\equiv R(r), \label{eq:def R}\\
\left(\frac{\mathrm{d}S^{(\theta)}}{\mathrm{d}\theta}\right)^2&=\frac{\kappa G^{(\theta)}-4L^2A^{(\theta)}+4E^2C^{(\theta)}+4ELB^{(\theta)}-K}{\mathcal{F}(\theta)}\nn\\&\equiv \Theta(\theta), \label{eq:def Theta}
\end{align}
where we have defined two compact functions $R(r)$ and $\Theta(\theta)$ to simplify the notation. Once the metric functions are known, these two functions are determined too. 
The function $S^{(r)}$ and $S^{(\theta)}$ therefore can be solved and the action \eqref{eq:the action of the free particle} becomes 
\begin{equation}
    S=-\frac{1}{2}\kappa\lambda-Et+L\phi+\int \pm_r\sqrt{R}\mathrm{d}r+\int\pm_\theta\sqrt{\Theta}\mathrm{d}\theta,
\end{equation}
where $\pm_r$ and $\pm_\theta$ are two signs introduced when taking the square roots in Eqs. \eqref{eq:def R} and \eqref{eq:def Theta}.
The motion equations for $r$ and $\theta$ coordinates then are found using
\begin{equation}
    \frac{\partial S}{\partial x^\mu}=P_\mu=g_{\mu\nu}\dot{x}^\nu
\end{equation}
to be 
\begin{align}
    \dot r=\frac{\pm_r\sqrt{R}}{D(r,\theta)} \label{eq:motion of r},\\
\dot{\theta}=\frac{\pm_\theta\sqrt{\Theta}}{F(r,\theta)}. \label{eq:motion of theta}
\end{align}

\subsection{Defining the deflection $\Delta\phi$ and $\Delta\theta$}

One of the main goals of this work is to find the deflection angles of the trajectory in the WDL. Denoting the source and the detector coordinates as $(r_s,\phi_s,\theta_s)$ and $(r_d,\phi_d,\theta_d)$ respectively, this goal is equivalent to finding 
\begin{align}
\Delta\phi\equiv\phi_d-\phi_s~\text{and}~ \Delta\theta\equiv\theta_d+\theta_s-\pi. \label{eq:dphidthetadef}
\end{align}
In the WDL, it is reasonable to assume that during the propagation of the signal, it only experiences one periapsis with radius $r_0\gg M$, where $M$ is the characteristic length scale of the spacetime, and one extreme value of the azimuth angle $\theta_m$. I.e., 
\begin{equation}
    \dot{r}|_{r=r_0}=0,~~ 
\dot{\theta}|_{\theta=\theta_m}=0. \label{eq:r0thmdef}
\end{equation}
The existence of such $\theta_m$ means it is either closer to $0$ or $\pi$ than both $\theta_s$ and $\theta_d$, and therefore we always have $|\cos\theta_m|>|\cos\theta_{s,d}|$.
From Eqs. \eqref{eq:motion of r} and \eqref{eq:motion of theta}, we see that the above can be inverted to find 
\begin{equation}
    r_0=R^{-1}(0),~~\theta_m=\Theta^{-1}(0).
\end{equation}
After substituting Eqs. \eqref{eq:def R} and \eqref{eq:def Theta}, this yields the more explicit relation
\begin{align}
&L=\frac{E(B^{r_0}+B^{\theta_m})+s_2\sqrt{\Xi}}{2(A^{r_0}+A^{\theta_m})},\label{eq:solved L}\\
&K=\frac{2E\left(A^{r_0}B^{\theta_m}-A^{\theta_m}B^{r_0}\right)\left[E\left(B^{r_0}+B^{\theta_m}\right)+s_2\sqrt{\Xi}\right]}{(A^{r_0}+A^{\theta_m})^2}\nn\\&+\frac{\kappa\left(A^{r_0}G^{\theta_m}-A^{\theta_m}G^{r_0}\right)+4E^2\left(A^{r_0}C^{\theta_m}-A^{\theta_m}C^{r_0}\right)}{A^{r_0}+A^{\theta_m}},\label{eq:solved GCC}
\end{align}
where
\begin{align}
&\Xi=(A^{r_0}+A^{\theta_m})\left[\kappa(G^{r_0}+G^{\theta_m})+4E^2(C^{r_0}+C^{\theta_m})\right]\nn\\
&~~~~+E^2(B^{r_0}+B^{\theta_m})^2,\nn\\
&X^{r_0}= X^{(r)}(r_0),\,X^{\theta_m}= X^{(\theta)}(\theta_m), ~X\in\{A,\,B,\,C,\,G\}\nonumber,
\end{align}
and $s_2=\pm 1$ is introduced when solving a quadratic equation. These relations connect the motion constants $(E,\,L,\,K)$ with $(r_0,\,\theta_m)$. 
In Sec. \ref{sec:The Perturbative Method}, we will use $(r_0,\,\theta_m)$ to replace $(L,\,K)$ since the latter are less intuitive and usually harder to measure in astronomy. For example, $r_0$ for the bending by the Sun can be approximated by the solar radius. 

To obtain the deflections $\Delta\phi$ and $\Delta\theta$, we first transform slightly the equations of motion \eqref{eq:firstintegralphi}, \eqref{eq:motion of r} and \eqref{eq:motion of theta} and show that they can be integrated. 
First of all, from Eqs. \eqref{eq:motion of r} and \eqref{eq:motion of theta}, one can find easily 
\begin{align}
   \mathrm{d}\lambda=\frac{D}{\pm_r\sqrt{R}}&\mathrm{d}r=\frac{F}{\pm_{\theta}\sqrt{\Theta}}\mathrm{d}\theta, \label{eq:dtau}
\end{align}
which after dividing $G(r,\theta)$ and using Eq. \eqref{eq:decompositionF} yields
\begin{align}
    \frac{1}{\pm_r\sqrt{R}\mathcal{D}}\mathrm{d}r&=\frac{1}{\pm_\theta\sqrt{\Theta}\mathcal{F}}\mathrm{d}\theta .\label{eq:dr dtheta}
\end{align}
On the other hand, substituting Eq. \eqref{eq:decompositionX} into Eq. \eqref{eq:firstintegralphi}, we have 
\begin{align}
\mathrm{d}\phi&=\frac{4LA^{(r)}-2EB^{(r)}+4LA^{(\theta)}-2EB^{(\theta)}}{G(r,\theta)}\mathrm{d}\lambda .\label{eq:dphi}
\end{align}
After using Eqs. \eqref{eq:decompositionF} and \eqref{eq:dr dtheta}, the $r$ and $\theta$ dependent parts in this equation are separated
\begin{align}
 \mathrm{d}\phi=
 \frac{4LA^{(r)}-2EB^{(r)}}{\pm_r\sqrt{R}\mathcal{D}}\mathrm{d}r+\frac{4LA^{(\theta)}-2EB^{(\theta)}}{\pm_{\theta}\sqrt{\Theta}\mathcal{F}}\mathrm{d}\theta .\label{eq:dphi2}
\end{align}
Integrating Eq. \eqref{eq:dphi2} we will directly obtain $\Delta\phi$
\begin{align}
\Delta\phi=&\left[\int^{r_s}_{r_0}+\int^{r_d}_{r_0}\right]\frac{4LA^{(r)}-2EB^{(r)}}{\sqrt{R}\mathcal{D}}\mathrm{d}r\nn\\&+s_1\left[\int^{\theta_s}_{\theta_m}+\int^{\theta_d}_{\theta_m}\right]\frac{4LA^{(\theta)}-2EB^{(\theta)}}{\sqrt{\Theta}\mathcal{F}}\mathrm{d}\theta. \label{eq:vip2}
\end{align}
Note that when integrating from $r_s$ to $r_0$ (or $r_0$ to $r_d$), the first term of Eq. \eqref{eq:dphi2} is expected to have $\pm_r=-1$ (or $\pm_r=+1$). And when integrating from $\theta_s$ to $\theta_m$ (or $\theta_m$ to $\theta_d$), $\pm_{\theta}=-1$ (or $\pm_{\theta}=+1$) if $\theta_m$ is a minimum or $\pm_{\theta}=+1$ (or $\pm_{\theta}=-1$) if $\theta_m$ is a maximum.  These sign values caused the extra $s_1=\mathrm{sign}(\cos(\theta_m))$ sign in front of the second integral in Eq. \eqref{eq:vip2}.
Similarly, integrating Eq. \eqref{eq:dr dtheta} we will obtain the following relation between initial and final $\theta$ coordinates 
\begin{equation}
\left[\int^{r_s}_{r_0}+\int^{r_d}_{r_0}\right]\frac{1}{\sqrt{R}\mathcal{D}}\mathrm{d}r=s_1\left[\int^{\theta_s}_{\theta_m}+\int^{\theta_d}_{\theta_m}\right]\frac{1}{\sqrt{\Theta}\mathcal{F}}\mathrm{d}\theta. \label{eq:vip1}
\end{equation}
This relation allows us to solve $\theta_d$ once $\theta_s$, the spacetime, and other kinetic variables are known. Therefore from this, we can find the deflection in the \mtheta direction as defined in Eq. \eqref{eq:dphidthetadef}.

\section{Perturbative Method and Deflections\label{sec:The Perturbative Method}}

The integrations \eqref{eq:vip2} and  \eqref{eq:vip1} which solve the deflection angles however, usually can not be carried out to obtain closed analytical forms. 
Therefore in this section, we will develop the perturbative method to approximate these integrals and then obtain the deflection. 

\subsection{The Perturbative Method}

The main idea of the perturbative method is selecting appropriate expansion parameter(s) and expanding the integrands into simpler series so that the integrations become doable. In the WDL, there is a naturally small parameter $1/r_0$ suitable for this purpose. When expanding the integrands in Eqs. \eqref{eq:vip2} and \eqref{eq:vip1}, one can also anticipate that the expansion coefficients will explicitly depend on the asymptotic behavior of the metric functions. 
After a short survey of the applicable spacetime metrics of our method, we found the following expansions can be assumed for the functions $X^{(\mu)} ~(\mu=r,\,\theta)$ and $\mathcal{D}(r)$ and $\mathcal{F}(\theta)$ 
\begin{subequations}
\label{eq:originalcoeffcients1}
\begin{align}
    &A^{(r)}=\sum^{\infty}_{n=2}\frac{a_n}{r^n},&&A^{(\theta)}=\frac{1}{4\sin^2\theta}, \\ 
    &B^{(r)}=\sum^{\infty}_{n=2}\frac{b_n}{r^{n-1}}, &&B^{(\theta)}=0,\\ 
    &C^{(r)}=\sum^{\infty}_{n=0}\frac{c_n}{r^{n-2}},&&C^{(\theta)}=-\frac{a^2\sin^2\theta}{4},\\ 
    &\mathcal{D}(r)=\sum^{\infty}_{n=0}\frac{d_n}{r^{n-2}},&&\mathcal{F}(\theta)=1,\\  
    &G^{(r)}=\sum^{\infty}_{n=0}\frac{g_n}{r^{n-2}}, &&G^{(\theta)}=a^2\cos^2\theta,
\end{align}
\end{subequations}
where the constant $a$ can be interpreted as the spacetime spin, and without losing any generality, we can always assume $a\geq 0$. Other coefficients $a_n,\,b_n,\,c_n,\,d_n,\,g_n$ can be determined once the metric functions are known. Note that for the $\theta$ functions of the above form, the relation \eqref{eq:r0thmdef} between $\theta_m$ and other parameters becomes very explicit as
\begin{align}
&a^2\left(E^2+\kappa\right)c_m^4-\left(K+2a^2E^2+\kappa a^2\right)c_m^2\nn\\&+a^2E^2+K+L^2=0
 \end{align}
and in principle, we can solve $\theta_m$ in terms of other parameters if needed.
Substituting the above series and using the following simple changes of variables 
\begin{align}
p\equiv \frac{r_0}{r},~c\equiv \cos\theta,~s\equiv \sin\theta
\end{align} in the integrals of Eqs. \eqref{eq:vip2} and \eqref{eq:vip1},
they can be further expanded with $1/r_0$ as the small parameter into the following series forms
\begin{align}
&\Delta\phi=\bigg[\int_{1}^{p_s}+\int^{p_d}_{1}\bigg]\sum_{i=2}^{\infty}n_{r,i}(p)\left(\frac{1}{r_0}\right)^i \mathrm{d}p\nn\\
&~~~~~~+s_1\bigg[\int^{c_s}_{c_m}+\int^{c_d}_{c_m}\bigg]\sum_{i=0}^{\infty}\frac{n_{\theta,i}(c)}{\sqrt{c_m^2-c^2}}\left(\frac{1}{r_0}\right)^i \mathrm{d}c,\label{eq:dphiexp}\\
&\bigg[\int_{1}^{p_s}+\int^{p_d}_{1}\bigg]\sum_{i=1}^{\infty}m_{r,i}(p)\left(\frac{1}{r_0}\right)^i \mathrm{d}p\nn\\
=&s_1\bigg[\int^{c_s}_{c_m}+\int^{c_d}_{c_m}\bigg]\sum_{i=1}^{\infty}\frac{m_{\theta,i}(c)}{\sqrt{c_m^2-c^2}}\left(\frac{1}{r_0}\right)^i \mathrm{d}c,
\label{eq:thetaint1}
\end{align}
where $c_{s,d,m},\,s_{s,d,m}$ and the small dimensionless quantities $p_{s,d}$ are defined as
\begin{align}
p_{s,d}= r_0/r_{s,d},\, c_{s,d,m}= \cos\theta_{s,d,m},\,s_{s,d,m}= \sin\theta_{s,d,m}. \nn
\end{align}
The coefficients $n_{r, i},\,n_{\theta, i},\,m_{r, i},\,m_{\theta, i}$ can be computed to any desired high order. Here we only list the first few orders of them
\begin{subequations}
\label{eq:expansion2}
\begin{align}
    n_{r,2}=&\frac{2p}{\sqrt{d_0(1-p^2)}}\bigg(\frac{ b_2 E}{\sqrt{\kappa g_0+4E^2c_0}}-2s_2s_ma_2p\bigg),\\ 
    n_{\theta,0}=&-\frac{s_2s_m}{1-c^2},\\ 
    n_{\theta,1}=&0,\\ 
    n_{\theta,2}=&-\frac{s_2s_ma^2(\kappa+E^2)}{2(\kappa g_0+4E^2c_0)},\\
        m_{r,1}=&-\frac{1}{\sqrt{(\kappa g_0+4E^2c_0)(1-p^2)d_0}},\\
    m_{r,2}=&\frac{(4E^2c_0+\kappa g_0)d_1p(1+p)+(4E^2c_1+\kappa g_1)d_0p}{2(1+p)\sqrt{(\kappa g_0+4E^2c_0)^3(1-p^2)d_0^3}},\\
    m_{\theta,1}=&-\frac{1}{\sqrt{\kappa g_0+4E^2c_0}},\\
    m_{\theta,2}=&\frac{\kappa g_1+4E^2c_1}{2(\kappa g_0+4E^2c_0)^{3/2}}.
\end{align}
\end{subequations}
Some higher-order terms are presented in Appendix \ref{sec:appendix1}.  
For the integrals over $p$ in Eqs. \eqref{eq:dphiexp} and \eqref{eq:thetaint1}, we can show that their integrands are always of the form $\text{polynomrial}(p)/\left(1-p^2\right)^{n+1/2}~(n=0,\,1,\,\cdots)$ and therefore integrable \cite{Jiang:2023yvo}. For the integral over $c$ in these equations, their integrands are always of the form $\text{polynomrial}(c)/\left[(1-c^2)^n \sqrt{c_m^2-c^2}\right]~(n=0,\,1)$ and therefore also integrable. 

The results of the integrations are of the form
\begin{align}
&\Delta\phi=\sum_{j=s,d}\sum^{\infty}_{i=2}N_{r,i}(p_j)\left(\frac{1}{r_0}\right)^i+\sum_{j=s,d}\sum^{\infty}_{i=0}N_{\theta,i}(c_j,c_m)\left(\frac{1}{r_0}\right)^i,\label{eq:expansion4}
\\
&\sum_{j=s,d}\sum^{\infty}_{i=1}M_{r,i}(p_j)\left(\frac{1}{r_0}\right)^i=\sum_{j=s,d}\sum^{\infty}_{i=1}M_{\theta,i}(c_j,c_m)\left(\frac{1}{r_0}\right)^i,\label{eq:expansion3}
\end{align}
where $N_{r,i},\,N_{\theta,i},\,M_{r,i},\,M_{\theta,i}$ are the corresponding integral results of the  $n_{r,i},\,n_{\theta,i},\,m_{r,i},\,m_{\theta,i}$ terms respectively. Again, the first few terms are 
\begin{subequations}
\label{eq:integralcoeffcients1}
\begin{align}
    N_{r,2}&=\frac{2s_2s_ma_2}{\sqrt{d_0}}\left[p_j\sqrt{1-p_j^2}+\cos^{-1}\left(p_j\right)\right]\nn\\&~~~~-\frac{2b_2E\sqrt{1-p_j^2}}{\sqrt{d_0(\kappa g_0+4E^2c_0)}},\\ 
    N_{\theta,0}&=\frac{s_2\pi}{2}-s_1s_2\tan^{-1}\left(\frac{c_js_m}{\sqrt{c_m^2-c_j^2}}\right),\\
    N_{\theta,1}&=0,\\
    N_{\theta,2}&=\frac{s_2s_ma^2(\kappa+E^2)}{2(\kappa g_0+4E^2c_0)}\cos^{-1}\left(\frac{c_j}{c_m}\right),\\
     M_{r,1}&=\frac{\cos^{-1}\left(p_j\right)}{\sqrt{(\kappa g_0+4E^2c_0)d_0}},\\ 
    M_{r,2}&=\frac{4E^2c_1d_0+\kappa g_1d_0}{2\left[\left(\kappa g_0+4E^2c_0\right)d_0\right]^{3/2}}\Bigg[\sqrt{\frac{1-p_j}{1+p_j}}-\cos^{-1}(p_j)\nn\\&~~~~-\frac{(\kappa g_0+4E^2c_0)d_1}{(\kappa g_1+4E^2c_1)d_0}\sqrt{1-p_j^2}\Bigg],\\ 
    M_{\theta,1}&=\frac{1}{\sqrt{\kappa g_0+4E^2c_0}}\cos^{-1}\left(\frac{c_j}{c_m}\right),\\
    M_{\theta,2}&=-\frac{(\kappa g_1+4E^2c_1)}{2(\kappa g_0+4E^2c_0)^{3/2}}\cos^{-1}\left(\frac{c_j}{c_m}\right).
\end{align}
\end{subequations}
and some higher-order results are given in Appendix \ref{sec:appendix1}.

\subsection{Deflection angles} 

We note that the deflection $\Delta\phi$ in Eq. \eqref{eq:expansion4} still contains the unknown $\theta_d$ in its second term coefficients $N_{\theta,i}$. On the other hand,   
Eq. \eqref{eq:expansion3} effectively establishes a relation between $\theta_d$ (or $\cos\theta_d$) and other parameters. Therefore to solve the deflections $\Delta\phi$ and  $\Delta\theta$, we will have to solve $\cos\theta_d$ from Eq. \eqref{eq:expansion3} first.
In the WDL,  $\cos\theta_d$ can also be expressed in the series form 
\begin{equation}
\cos\theta_d=\sum_{i=0}^{\infty}h_i\left(\frac{1}{r_0}\right)^i. \label{eq:cos theta d}
\end{equation}
where the coefficients $h_i$ are solvable from Eq. \eqref{eq:expansion3} using the method of undetermined coefficients. Here we only show the first two orders 
\begin{subequations}
\label{eq:h coeffcients}
\begin{align}
    h_0=&c_m\cos\left[\frac{1}{\sqrt{d_0}}\sum_{j=s,d}\cos^{-1}(p_j)-\cos^{-1}\left(\frac{c_s}{c_m}\right)\right],\\ 
    h_1=&-s_1\sqrt{c_m^2-h_0^2}\Bigg[\sqrt{\kappa g_0+4E^2c_0}\sum_{j=s,d}M_{r,2}(p_j)\nn\\&+\frac{\kappa g_1+4E^2c_1}{2(\kappa g_0+4E^2c_0)\sqrt{d_0}}\sum_{j=s,d}\cos^{-1}(p_j)\Bigg].
\end{align}  
\end{subequations}
and higher order ones are again in Appendix \ref{sec:appendix1}.

Substituting solution Eq. \eqref{eq:cos theta d} into Eq. \eqref{eq:expansion4} and carrying out the small $1/r_0$ expansion again, the deflection $\Delta\phi$ is found finally as 
\begin{equation}
    \Delta\phi=\sum_{j=s,d}\sum^{\infty}_{i=2}N_{r,i}(p_j)\left(\frac{1}{r_0}\right)^i+s_2\sum^{\infty}_{i=0}N'_{\theta,i}(c_s,c_m)\left(\frac{1}{r_0}\right)^i \label{eq:delta phi2}
\end{equation}
where the $N_{r,i}$ were unchanged as in Eq. \eqref{eq:integralcoeffcients1} and the first few $N^\prime_{\theta,i}$ as
\begin{subequations}
\begin{align}
    N'_{\theta,0}=&\pi-s_1\left[\tan^{-1}\left(\frac{c_ss_m}{\sqrt{c_m^2-c_s^2}}\right)+\tan^{-1}\left(\frac{h_0s_m}{\sqrt{c_m^2-h_0^2}}\right)\right],\\
    N'_{\theta,1}=&\frac{s_1s_mh_1}{(h_0^2-1)\sqrt{c_m^2-h_0^2}},\\
    N'_{\theta,2}=&\frac{s_1s_mh_0h_1^2(3h_0^2-2c_m^2-1)}{2(h_0^2-1)^2\left(c_m^2-h_0^2\right)^{3/2}}+\frac{s_1s_mh_2}{(h_0^2-1)\sqrt{c_m^2-h_0^2}}\nn\\&+\frac{s_ma^2(\kappa+E^2)}{2(\kappa g_0+4E^2c_0)}\left[\cos^{-1}\left(\frac{c_s}{c_m}\right)+\cos^{-1}\left(\frac{h_0}{c_m}\right)\right].
\end{align}
\end{subequations}
Inspecting the above results, one can discover that $s_2$ is just the sign of $\Delta\phi$ at the lowest order, which also means that $s_2=\pm1$ correspond to anticlockwise and clockwise motions respectively. Naturally, the following relation holds 
\begin{align}
    s_2=\mathrm{sign}(L).
\end{align}
In the infinite $r_{s,d}$ limit, we see clearly from the $N^\prime_{\theta,0}$ term that $\Delta\phi$ to the leading order equals $s_2\pi$. 

Similarly, substituting solution \eqref{eq:cos theta d} into Eq. \eqref{eq:dphidthetadef}, the deflection in $\theta$ direction becomes 
\begin{equation}
    \Delta \theta=\theta_s+\theta_d-\pi=\sum_{i=0}^{\infty}k_i\left(\frac{1}{r_0}\right)^i. \label{eq:delta theta2}
\end{equation}
where
\begin{subequations}
\begin{align}
    k_0=&\theta_s-\pi+\cos^{-1}(h_0),\\ 
    k_1=&-\frac{h_1}{\sqrt{1-h_0^2}},\\
    k_2=&\frac{2h_0^2h_2-h_0h_1^2-2h_2}{2(1-h_0^2)^{3/2}}.
\end{align}
\end{subequations}
Note that in the limit $r_{s,d}\to\infty$, $h_0$ approaches $-c_s$ and therefore $k_0$ approaches 0. 

Eqs. \eqref{eq:delta phi2} and \eqref{eq:delta theta2} are two important results of the work and a few comments are for them.  Firstly, these results apply to general SAS spacetimes that allow the existence of a GCC. This includes many well-known spacetimes such as the Kerr, KN, and all SSS ones, which can be obtained by setting all $b_n=0$ for $n\geq2$. 
Secondly, they work for both light rays (setting $\kappa=0$) and timelike particles (setting $\kappa=-1$). Indeed we can show that the $E\to\infty$ limits of these results for timlike signals equal exactly their values for null rays. 
Thirdly, these deflections also take into account the finite distance effect of the source and detector. This effect could be important when studying the GL effect. Setting $p_{s,d}$ to zero, the infinite distance version of the deflections can be obtained. 
Fourthly, these deflections work for both non-equatorial and equatorial trajectories. For the former, setting $\theta_{s,d}\to \pi/2$, we have verified that the deflection angle $\Delta\phi$ reduces to its value on the equatorial plane in the Kerr spacetime \cite{Huang:2020trl}. 
For the latter case, these formulas do not rely on any near-equatorial plane approximation, i.e., $\theta_s$ and $\theta_d$ can be far from $\pi/2$. Last but not least, the deflections \eqref{eq:delta phi2} and \eqref{eq:delta theta2} can be further expanded around small $p_{s,d}$ if the source and detector are far away, i.e., $r_{s,d}\gg r_0$, and a dual series form will be obtained. Such a form will be more appealing from the application point of view and we will only do this in Sec. \ref{sec:Applications}. 

\section{Gravitational lensing\label{sec:Gravitational Lensing Effects}}

\begin{figure}[htp!]
   \centering
   \includegraphics[width=0.45\textwidth]{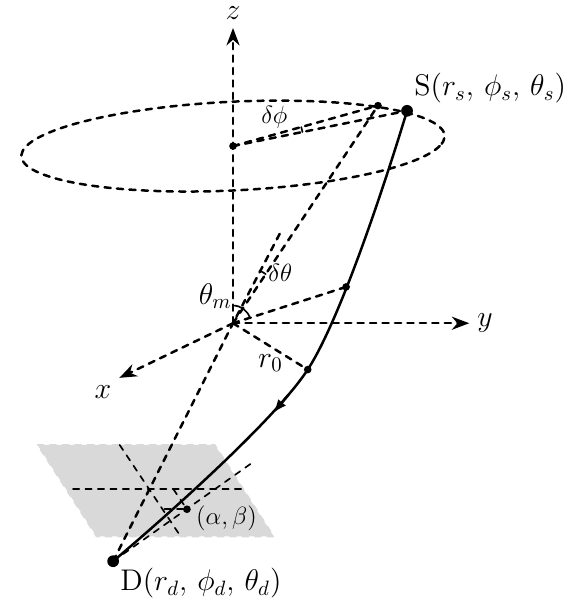}\\
   \caption{Schematic diagram for one trajectory from the source at S$(r_s,\,\phi_s,\,\theta_s)$ to the detector at D$(r_d,\,\phi_d,\,\theta_d)$. The $r_0$ and $\theta_m$ mark the minimal radius and extreme $\theta$ points on the trajectory. The $(\alpha,\,\beta)$ marks the apparent angle formed by this trajectory on the celestial sphere of the observer.}
   \label{fig:schm}
\end{figure}

As we've got the deflection angles for arbitrary inclination angles of the signals, we can study the GL in such SAS spacetime in the off-equatorial plane. Since our deflection angles \eqref{eq:delta phi2} and \eqref{eq:delta theta2} contain the finite distance effect, we can naturally establish the following GL equations
\begin{subequations}
\label{eq:gleq}
\begin{align}
&\delta\phi= \Delta\phi-s_2\pi,\label{eq:def delta phi}\\
&\delta\theta=\Delta\theta. \label{eq:def delta theta}
\end{align}
\end{subequations}
Here $\delta\phi$ and $\delta\theta$ are the two small angles characterizing the angular position of the source relative to the detector-lens axis in the spherical coordinates as shown in the schematic diagram in Fig. \ref{fig:schm}. Once $r_{s,d},\,\theta_s$ are fixed, then substituting Eqs. \eqref{eq:delta phi2} and \eqref{eq:delta theta2} into Eq. \eqref{eq:gleq} will allow us to solve the minimal radius and extreme azimuth angle $(r_0,\,\theta_m)$ for each pair of $\delta\phi,\,\delta\theta$. Unfortunately, Eq. \eqref{eq:gleq} are high-order polynomials of $r_0$ and more complicated functions of $\theta_m$, which usually can not be solved analytically. This is particularly so when the effects of higher-order parameters are sought. Therefore when solving them, most of the time we will use the numerical method. 

It is found that in general, there will be two sets of $(r_0,\,\theta_m)$ that allow the signal to reach the detector. Most of the time when the deflection $(\delta\theta,\,\delta\phi)$ is not small, these two solutions will have opposite orbital rotation direction $s_2$ and the $s_1=\mathrm{sign}(\cos(\theta_m))$ equals $s_2$ for each solution. Only when the source, lens and detector are very aligned (deflection less than $10^{-6\prime\prime}$ in the Sgr A* situation considered in Sec. \ref{sec:kn}) and the spacetime spin is large, there could exist exceptions (see also Ref. \cite{Jiang:2023yvo}). Therefore we will label these two solutions as $(r_{0+},\,\theta_{m+})$ and $(r_{0-},\,\theta_{m-})$ to represent the prograde and retrograde rotating signals respectively. In this paper, by {\it prograde} and {\it retrograde}, we always and only mean that the trajectories are rotating anticlockwise and clockwise around the $+\hat{z}$ directions respectively. No retrolensing is involved because we only discuss the weak deflection cases in this work.

To link the solved $(r_0,\,\theta_m)$ to the observables of the GL however, we still need to work out the formula for the apparent angles of the lensed images. For a static observer in the spacetime with metric \eqref{eq:general metric}, the associated tetrad $(e_a)^\mu$ takes the form
\begin{subequations}
    \begin{align}
e_0=&\frac{1}{\sqrt{A}}\frac{\partial}{\partial t}\equiv Z,\\
e_1=&-\sqrt{\frac{B^2}{AB^2+4A^2C}}\left(\frac{\partial}{\partial t}+\frac{2A}{B}\frac{\partial}{\partial \phi}\right)\equiv \hat{\phi},\\
e_2=&\frac{1}{\sqrt{D}} \frac{\partial}{\partial r} \equiv\hat{r},\\
e_3=&\frac{1}{\sqrt{F}} \frac{\partial}{\partial \theta} \equiv\hat{\theta},
\end{align}
\end{subequations}
where $Z$ is the four-velocity of the observer.
Then for a signal with four-velocity
$u^\mu=(\dot{t},\,\dot{\phi},\,\dot{r},\,\dot{\theta})$, its projection using the projection operator 
$h^\mu_{\ \nu}=\delta^\mu_{\ \nu}+Z^\mu Z_\nu$
into the tetrad frame yields the vector
\begin{align}
\tilde{u}^\mu=& h^\mu_{\ \nu}u^\nu
=
-\frac{B\dot\phi}{2}\sqrt{\frac{B^2+4AC}{AB^2}}\hat{\phi}+\sqrt{D}\dot{r}\hat{r}+\sqrt{F}\dot{\theta}\hat{\theta}.
\end{align}
Here the $(\dot{r},\,\dot{\theta},\,\dot{\phi})$ are linked to $(L,\,K)$ through Eqs. \eqref{eq:firstintegralphi}, \eqref{eq:motion of r} and  \eqref{eq:motion of theta} and then to the $(r_{0\pm},\,\theta_{m\pm})$ through Eqs. \eqref{eq:solved L} and \eqref{eq:solved GCC}.
Then the apparent angle $\gamma_\pm$ measured by this observer against the detector's radial direction $\hat{r}_d$ and $\alpha_\pm$ against the $\hat{r}_d\hat{\theta}_d$ plane and $\beta_\pm$ against the $\hat{r}_d\hat{\phi}_d$ plane are respectively 
\begin{subequations}
\label{eq:apparentangles}
\begin{align}
    \gamma_\pm&=\cos^{-1}\frac{(\tilde{u},\hat{r})}{|\tilde{u}||\hat{r}|}\nn\\
    &=\cos^{-1}\lb\dot{r}\sqrt{\frac{D}{\frac{B^2+4AC}{4A}\dot{\phi}^2+D\dot{r}^2+F\dot{\theta}^2}}\rb\Bigg|_d,\label{eq:apparent angle gamma}\\
    \alpha_\pm&=\frac{\pi}{2}-\cos^{-1}\frac{(\tilde{u},\hat{\phi})}{|\tilde{u}||\hat{\phi}|}\nn\\
    &=-\sin^{-1}\lb\frac{B\dot{\phi}}{2A}\sqrt{\frac{(AB^2+4A^2C)/B^2}{\frac{B^2+4AC}{4A}\dot{\phi}^2+D\dot{r}^2+F\dot{\theta}^2}}\rb\Bigg|_d,\label{eq:apparent angle alpha}\\
    \beta_\pm&=\frac{\pi}{2}-\cos^{-1}\frac{(\tilde{u},\hat{\theta})}{|\tilde{u}||\hat{\theta}|}\nn\\
    &=\sin^{-1}\lb\dot{\theta}\sqrt{\frac{F}{\frac{B^2+4AC}{4A}\dot{\phi}^2+D\dot{r}^2+F\dot{\theta}^2}}\rb\Bigg|_d,\label{eq:apparent angle beta}
\end{align}
\end{subequations}
where the subscript $|_d$ means all coordinates should be evaluated at the detector location. These equations are valid for all trajectories in the SAS spacetimes including those that are bent strongly. In the large $r_d$ limit, then $\gamma_\pm^2\approx \alpha_\pm^2+\beta_\pm^2$ and therefore we only need to concentrate on two of them, which are conventionally chosen as $(\alpha_\pm,\,\beta_\pm)$. 

\section{Application to particular spacetimes\label{sec:Applications}}

In this section, we will apply the general method and results in Secs. \ref{sec:The Perturbative Method}  and \ref{sec:Gravitational Lensing Effects} to some known SAS spacetimes to examine the validity of the results and significance of the spacetime parameters on the off-equatorial deflection and GL. We will focus on deflections $\Delta\phi,\,\Delta\theta$, the $r_0,\theta_m$ and apparent angles $\alpha_\pm,\,\beta_\pm$.

\subsection{Deflections and GL in  KN spacetime}\label{sec:kn}

For KN BH, the deflection in its equatorial plane has been considered using analytical methods repeatedly and the (quasi-)equatorial motion was studied multiple times too \cite{Hsieh:2021scb,Kraniotis:2014paa,Ghosh:2022mka,He:2021rrz,Hsiao:2019ohy}, let alone numerical solution of the trajectories. However, to our best knowledge, the perturbative study of the general off-equatorial deflection has not been done yet.

The metric of KN is given by
\begin{align}
    \mathrm{d}s^2=&-\frac{\Sigma_{\mathrm{KN}}-2Mr+Q^2}{\Sigma_{\mathrm{KN}}}\mathrm{d}t^2-\frac{2a(2Mr-Q^2)\sin^2\theta}{\Sigma_{\mathrm{KN}}} \mathrm{d}t\mathrm{d}\phi\nn\\&+\frac{\left[\left(r^2+a^2\right)^2-\Delta_{\mathrm{KN}} a^2\sin^2\theta\right]\sin^2\theta}{\Sigma_{\mathrm{KN}}} \mathrm{d}\phi^2\nn\\&+\frac{\Sigma_{\mathrm{KN}}}{\Delta_{\mathrm{KN}}}\mathrm{d}r^2+\Sigma_{\mathrm{KN}} \mathrm{d}\theta^2, \label{eq:knmetric}
\end{align}
where
\begin{align}
    &\Sigma_{\mathrm{KN}}= r^2+a^2\cos^2\theta, \nn\\
    &\Delta_{\mathrm{KN}}= r^2-2Mr+a^2+Q^2,\nn
\end{align} 
and $M,\, Q,\,a=J/M$ are respectively the mass, charge and spin angular momentum per unit mass of the spacetime. In studying the trajectories, one can always choose $a\geq0$ if the motion is allowed to go both clock- and anticlock-wise. 
To use the method and results developed in Secs. \ref{sec:The Perturbative Method}  and \ref{sec:Gravitational Lensing Effects}, the first thing to check is whether the metric \eqref{eq:knmetric} satisfies the separation requirements \eqref{eq:originalcoeffcients1}. Substituting the metric into these equations, it is easy to find that the separation can be done and the $r$-dependent functions are 
\begin{subequations}
\label{eq:KN expansioncoeffcients}
\begin{align}
    A^{(r)}_{\mathrm{KN}}=&-\frac{a^2}{4\Delta_{\mathrm{KN}}}=-\frac{a^2}{4r^2}-\frac{Ma^2}{2r^3}+\mathcal{O}(r)^{-4},\\
    B^{(r)}_{\mathrm{KN}}=&\frac{aQ^2-Mar}{\Delta_{\mathrm{KN}}}\nn\\
    =&-\frac{Ma}{r}+\frac{-2M^2a+aQ^2}{r^2}+\mathcal{O}(r)^{-3},\\
    C^{(r)}_{\mathrm{KN}}=&\frac{\left(r^2+a^2\right)^2}{4\Delta_{\mathrm{KN}}}\nn\\
    =&\frac{r^2}{4}+\frac{Mr}{2}+\frac{a^2+4M^2-Q^2}{4}+\mathcal{O}(r)^{-1},\\
    \mathcal{D}(r)_{\mathrm{KN}}=&r^2-2Mr+a^2+Q^2,\\
    G^{(r)}_{\mathrm{KN}}=&r^2
\end{align}
\end{subequations}
and the $X^{(\theta)}~(X\in\{A,\,B,\,C,\,\mathcal{F},\,G\})$ are exactly as given in Eq. \eqref{eq:originalcoeffcients1}. This guarantees the existence of the GCC, as was well-known prior, and the applicability of the results of deflection angles. 

Substituting the coefficients in Eqs. \eqref{eq:KN expansioncoeffcients} directly into Eq. 
\eqref{eq:delta phi2} and \eqref{eq:delta theta2} and carrying out the small $p_{s,d}$ expansion, the deflection angles in KN spacetime are found as dual series of $M/r_0$ and $p_{s,d}$
\begin{widetext}
\begin{subequations}
\label{eq:dkn}
\begin{align}
    \Delta\phi_{\mathrm{KN}}=&s_2\pi+\frac{4\hat{a}M^2}{vr_0^2}-\frac{8s_m^2\hat{a}M^2}{s_s^2vr_0^2}+\frac{s_2s_m}{s_s^2}\left[\frac{2(1+v^2)M}{v^2r_0}-\left(p_s+p_d\right)-\frac{\zeta_{\mathrm{KN}}M^2}{4v^4r_0^2}-\frac{M\left(p_s+p_d\right)}{v^2r_0}\right]\nn\\&+\frac{s_1s_2s_mc_s\sqrt{c_m^2-c_s^2}}{s_s^4}\left[\left(p_s+p_d\right)-\frac{2(1+v^2)M}{v^2r_0}\right]^2+\mathcal{O} (\epsilon)^3,\label{eq:delta phi KN}\\
    \Delta\theta_{\mathrm{KN}}=&\frac{s_1\sqrt{c_m^2-c_s^2}}{s_s}\Bigg[\frac{2(1+v^2)M}{v^2r_0}-\left(p_s+p_d\right)-\frac{\left(\zeta_{\mathrm{KN}}+32s_2s_mv^3\hat{a}\right)M^2}{4v^4r_0^2}-\frac{M\left(p_s+p_d\right)}{v^2r_0}\Bigg]\nn\\&-\frac{c_ss_m^2}{2s_s^3}\left[\left(p_s+p_d\right)-\frac{2(1+v^2)M}{v^2r_0}\right]^2+\mathcal{O} \left(\epsilon\right)^3,\label{eq:delta theta KN}
\end{align}
\end{subequations}
\end{widetext}
where the infinitesimal $\epsilon$ represents either the $M/r_0$ or $p_{s,d}$, and
\begin{align}
\zeta_{\mathrm{KN}}=8+8v^2-12\pi v^2-3\pi v^4+\pi v^2(2+v^2)\hatcq^2,\label{zetaKN}
\end{align}
$\hat{a}\equiv a/M,\,\hat{Q}\equiv Q/M,$  and $E$ has been replaced by the asymptotic velocity $v$ through Eq. \eqref{eq:evrel}. There are a few limits that we can take for these deflections. Setting $v=1$ or $p_{s,d}=0$, they reduce respectively to deflections of light rays or deflections from infinity to infinity. A more unusual limit is to set $\hata=0$, which pushes these deflections to their values of the signal in an RN spacetime but with arbitrary incoming direction. Setting $\hatcq=0$, they agree with Eqs.  (32) and (39) of Ref. \cite{Jiang:2023yvo}. In the infinite distance and equatorial limit, $p_s,\,p_d,\,c_m,\,c_s\to 0$ and $s_m,\,s_s\to 1$, and we have checked that $\Delta\phi_{\mathrm{KN}}$ in this limit agrees with Eq. (83) of Ref. \cite{Jia:2020xbc}. 

Both Eqs. \eqref{eq:delta phi KN} and \eqref{eq:delta theta KN} illustrate various effects of the non-equatorial motion and spacetime parameters. 
For the deflection $\Delta\phi_{\mathrm{KN}}$, the following observations can be made. 
Firstly we observe that the non-equatorial effect manifests in two ways. The first is through the terms proportional to $c_s\sqrt{c_m^2-c_s^2}$. These terms will vanish in the equatorial limit and therefore we call them {\it non-equatorial} terms. The second way of the non-equatorial effect is through the factor $s_m^n/s_s^2$ of the other terms, which will be called the {\it equatorial} terms. If the trajectory was in the equatorial plane, these factors would all be one. Therefore these factors effectively adjusted the contribution of equatorial terms to the deflection. 
The second comment concerns the effects of the spacetime charge and spin. As in the case of equatorial motion, $\hatcq$ and $\hata$ start to appear in the equatorial terms from the second order only, i.e., the $(M/r_0)^2,\,(r_0/r_{s,d})^2$ or $(M/r_0)(r_0/r_{s,d})$ terms. While in non-equatorial terms, they start to appear from the third order.

The terms of the deflection $\Delta\theta$ in Eq. \eqref{eq:delta theta KN} are either proportional to $\sqrt{c_m^2-c_s^2}$ or $c_s$, which both approach zero in the equatorial limit. At the leading orders of both $(M/r_0)$ and $p_{s,d}$, the terms are proportional to $\sqrt{c_m^2-c_s^2}$. For the effect of both $\hatcq$ and $\hata$, they both appear from the second order, which is similar to the equatorial terms in $\Delta\phi_{\mathrm{KN}}$. 
In both $\Delta\phi_{\mathrm{KN}}$ and 
$\Delta\theta_{\mathrm{KN}}$, the sign of $\hatcq$ does not matter, as expected since the signal is neutral.

\begin{figure}[htp!]
   \centering
    \includegraphics[width=0.225\textwidth]{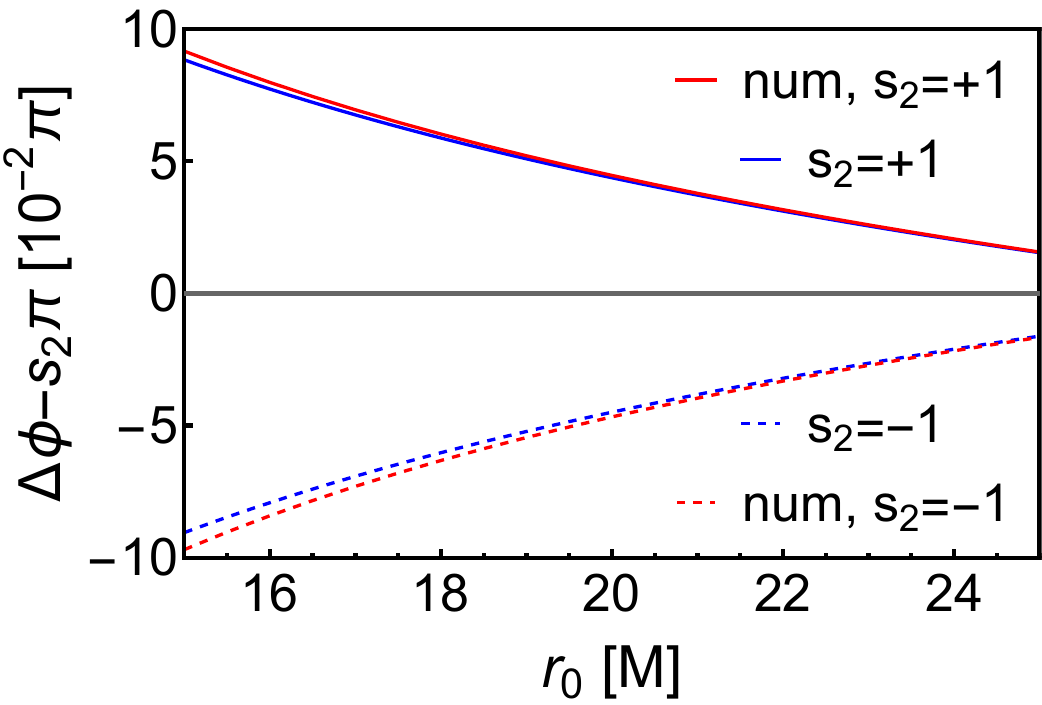}
   \includegraphics[width=0.225\textwidth]{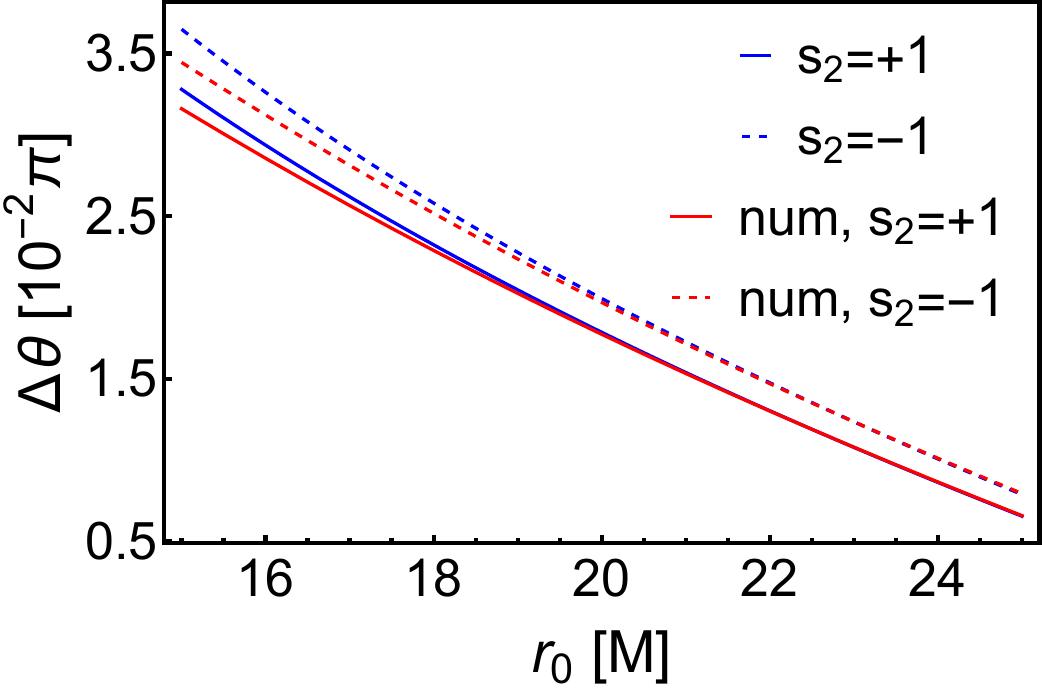}\\
   (a)\hspace{3.5cm}(b)\\
   \includegraphics[width=0.225\textwidth]{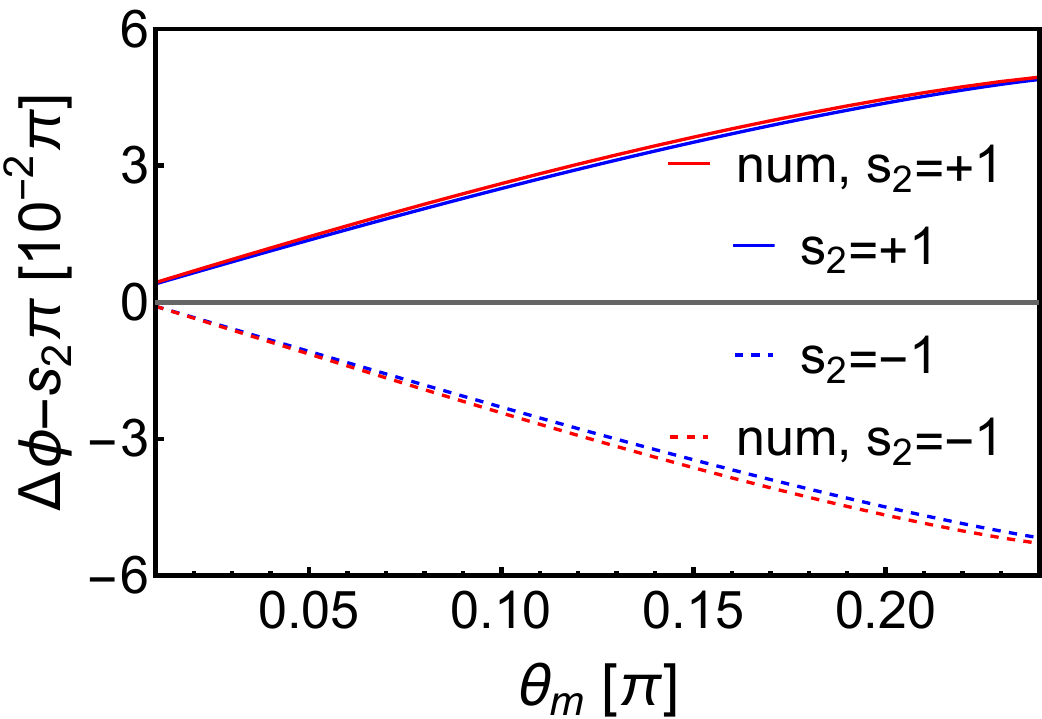}
   \includegraphics[width=0.225\textwidth]{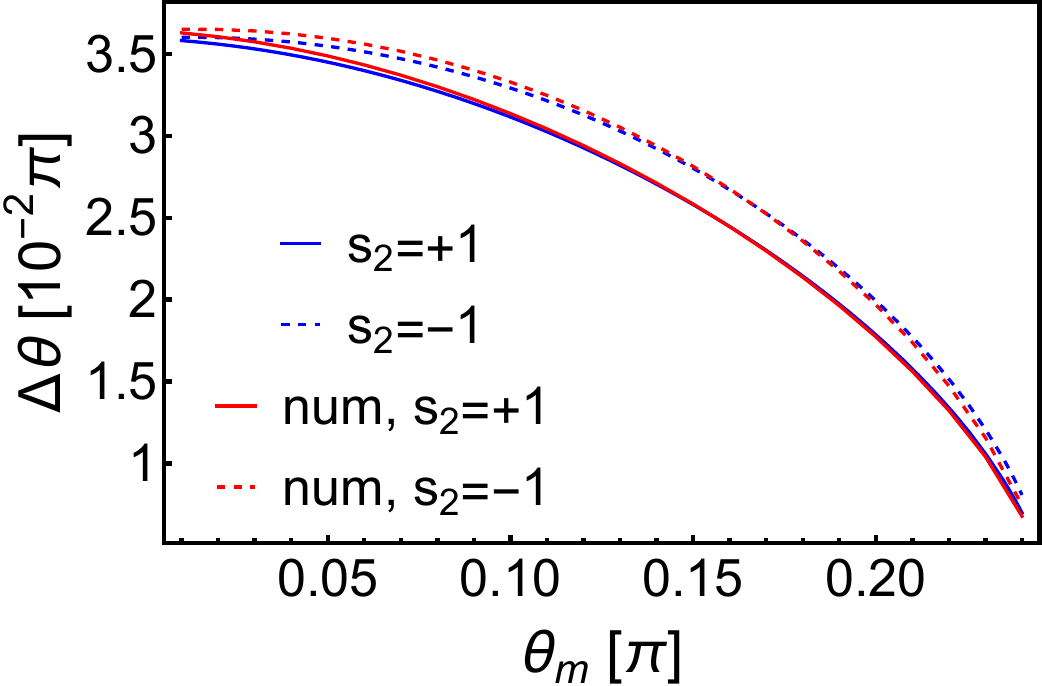}\\
   (c)\hspace{3.5cm}(d)\\
   \includegraphics[width=0.45\textwidth]{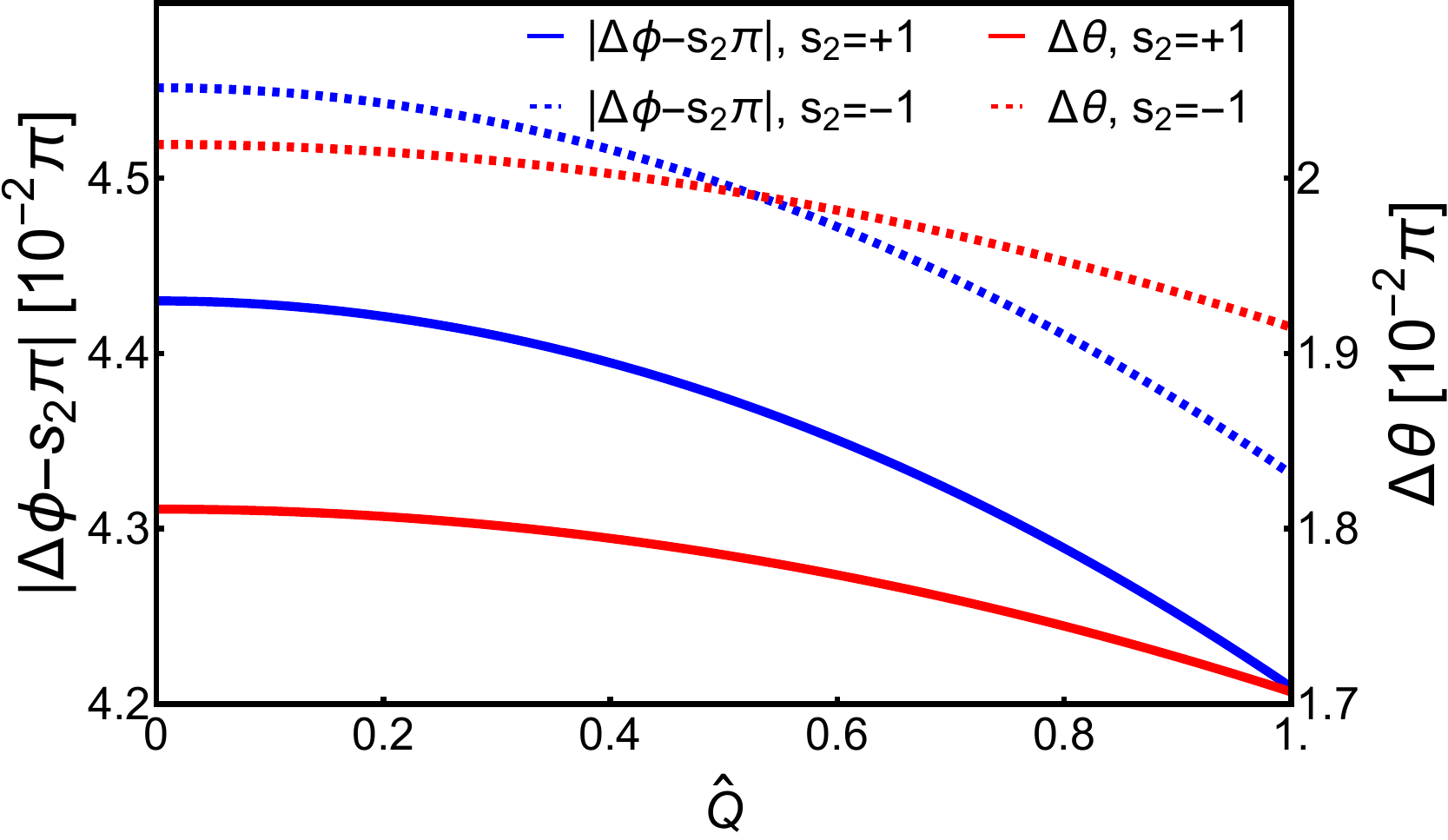}\\(e)\\
   \caption{The dependences of $\Delta\phi$ and $\Delta\theta$ on $r_0,\,\theta_m$ and $\hat{Q}$ in KN spacetime. $v=1,\,r_s=r_d=400M,\,\hat{a}=1/2,\,\theta_s=\pi/4$. In (a)(b), $\hatcq=1/2$, $\theta_m=\pi/5$. In (c)(d), $\hatcq=1/2,\,r_0=20M$. In (e), $r_0=20M,\,\theta_m=\pi/5$. All red lines represent numerical results.  }
   \label{fig:delta phi theta_KN}
\end{figure}

To check the validity of these deflections \eqref{eq:delta phi KN} and \eqref{eq:delta theta KN}, in Fig. \ref{fig:delta phi theta_KN} we compare them with their corresponding numerically integrated values. In this plot, we choose relatively small $r_0$ so that the deflections are appreciable to tell the effects of various parameters. It is seen from Fig. \ref{fig:delta phi theta_KN} (a) and (b) that as $r_0$ increases, both $|\Delta\phi_{\mathrm{KN}}|$ and $|\Delta\theta_{\mathrm{KN}}|$ decrease monotonically. The analytical results approach the numerical value more closely as $r_0$ increases too, which is expected because both deflections are series of $(M/r_0)$. From Fig. \ref{fig:delta phi theta_KN} (c) and (d), we see that as $\theta_m$ decreases, the deflection in $\theta$ direction increases while that in the $\phi$ direction decreases. This is intuitively consistent with the physical expectation because the decreasing of $\theta_m$ corresponds to the motion of the trajectory asymptotic line towards the $z$ axis above the equatorial plane. Fig. \ref{fig:delta phi theta_KN} (e) shows the effect of $\hatcq$ on the deflections. Previously it was known from the equatorial plane case that the $|\Delta\phi_{\mathrm{KN}}|$ would decrease as $\hatcq$ increase \cite{Hsiao:2019ohy}, which is still observed here for the off-equatorial motion. We also note that this is even true for $\Delta\theta_{\mathrm{KN}}$, which shows the spherical nature of the effect of $\hatcq$ on the deflections. Lastly, the effect of the spacetime spin $\hata$ on these deflections was studied in Ref. \cite{Jiang:2023yvo} in the Kerr spacetime and we found that qualitatively that effect is not changed: larger $\hata$ increases (or decreases) the $\delta\phi$ of the prograde (or retrograde) signal. 

With the correctness of the series results confirmed, we can now solve the lensing equations \eqref{eq:gleq} with $\Delta\theta$ and $\Delta\phi$ given by Eqs. \eqref{eq:delta theta KN} and \eqref{eq:delta phi KN} to obtain the $(r_0,\,\theta_m)$ for a fixed pair $(\delta\phi,\,\delta\theta)$ which characterizes the angular deflection of the source against the lens-observer axis. For the non-equatorial motion in a SAS spacetime, the source's azimuth angle $\theta_s$ also becomes important. Since these equations are high-order polynomials if the effects of $\hatcq$ and $\hata$ are taken into account, we have only solved them numerically. We used the Sgr A* BH as the lens and set $r_d=r_s=8.34~\mathrm{kpc}$ and varied $(\delta\phi,\,\delta\theta,\,\theta_s)$. We found that qualitatively the effect of $\delta\phi,\,\delta\theta,\,\theta_s$ and $\hata$ are similar to their effects in the Kerr case studied in Ref. \cite{Jiang:2023yvo}. For this reason, we will not show these figures here. Rather, we only mention that a larger positive $\hata$ decreases (or increases) the $r_0$ of a counterclockwise (or clockwise) rotating trajectory, and therefore the trajectory is pulled towards (or pushed away from) the $z$ axis. The $\theta_m$ will change accordingly: a larger positive $\hata$ will increase $|\cos\theta_m|$ of both counterclockwise and clockwise rotating orbits. For the charge $\hatcq$, its deviation from zero was found to decrease $r_0$  for all $\hata$ and orbit rotation directions and increases $|\cos\theta_m|$ very weakly. 

Finally, substituting the solved $(r_0,\,\theta_m)$ together with the initial parameters $(\delta\theta,\,\theta_s,\,r_d)$ into Eqs. \eqref{eq:apparent angle alpha} and \eqref{eq:apparent angle beta}, we can obtain the apparent positions of the two images on the celestial sphere of the detector 
\begin{subequations}
\begin{align}
&\alpha_{\pm\mathrm{KN}}\nn\\
=
&\sin^{-1}\frac{L_\pm\left(\Delta_{\mathrm{KN}}-a^2\sin^2\theta\right)+a\sin^2\theta E\left(2Mr-Q^2\right)}{\sin\theta\sqrt{\Delta_{\mathrm{KN}}\Sigma_{\mathrm{KN}}\left[\Sigma_{\mathrm{KN}} E^2+\kappa\left(\Sigma_{\mathrm{KN}}-2M r+Q^2\right)\right]}}\Bigg|_d,\\
&\beta_{\pm \mathrm{KN}}\nn\\
=&s_1\sin^{-1}\sqrt{\frac{\Theta_\pm\left(\Sigma_{\mathrm{KN}}-2M r+Q^2\right)}{\Sigma_{\mathrm{KN}}\left[\Sigma_{\mathrm{KN}} E^2+\kappa\left(\Sigma_{\mathrm{KN}}-2M r+Q^2\right)\right]}}\Bigg|_d,
\end{align}
\end{subequations}
where $\Theta_\pm$ was defined in Eq. \eqref{eq:def Theta} and takes the form
\begin{align}
\Theta_\pm=\kappa a^2\cos^2\theta-L_\pm^2\csc^2\theta-a^2E^2\sin^2\theta-K_\pm \label{eq:thetakn}
\end{align}
in KN spacetime. The $L_\pm,\,K_\pm$ can be fixed by $(r_{0\pm},\,\theta_{m\pm})$ through Eqs. \eqref{eq:solved L} and \eqref{eq:solved GCC}. These apparent angles are consistent with the results in Ref. \cite{Jiang:2023yvo}. 
For the null particle in the WDL, we can make the following power series approximation for them by taking $M/r_0$ and $M/r_d$ as small quantities
\begin{subequations}
\label{eq:aaKNapp}
\begin{align}    \alpha_{\pm \mathrm{KN}}\simeq&\frac{s_2s_{m\pm}}{s_d\hat{r}_d}\left(\hat{r}_{0\pm}+1+\frac{3+\hat{a}^2- \hat{Q}^2-4s_2s_{m\pm}\hat{a}}{2\hat{r}_{0\pm}}\right),
    \label{eq:alphaKNappro}\\
    \beta_{\pm \mathrm{KN}}\simeq&\frac{s_1\sqrt{s_d^2-s_{m\pm}^2}}{s_d\hat{r}_d}\left(\hat{r}_{0\pm}+1+\frac{3+\hat{a}^2 c_d^2- \hat{Q}^2-4s_2s_{m\pm}\hat{a}}{2\hat{r}_{0\pm}}\right),
    \label{eq:betaKNappro}
\end{align}
\end{subequations}
where and henthforce $\hat{r}_{0\pm}\equiv r_{0\pm}/M,~\hat{r}_d\equiv r_d/M,~s_{m\pm}=\sin\theta_{m\pm}$. When we set $\hatcq=0$, they agree with Ref. \cite{Jiang:2023yvo}. When $\hatcq\neq0$ however, its effect does not only appear from the $Q^2/(r_{0\pm}r_d)$ order as one might think superficially from Eq. \eqref{eq:aaKNapp}. Indeed, $\hatcq$ affects $\hatr_{0\pm}$ by an amount similar to the size of $\hatcq$ itself and therefore its influence on the image apparent angles is at the $Q/r_d$ order, i.e., one order lower than what appeared in Eq. \eqref{eq:aaKNapp}.  The off-equatorial effect influences $\gamma_{\pm\mathrm{KN}}$ from the second order because at the leading order the total apparent angle $\gamma_{\pm\mathrm{KN}}\approx\sqrt{\alpha_{\pm\mathrm{KN}}^2+\beta_{\pm\mathrm{KN}}^2}$ would not be tuned by the factor $s_{m\pm}/s_d$ or $\sqrt{s_d^2-s_{m\pm}^2}/s_d$ in front of $\alpha_{\pm\mathrm{KN}}$ and $\beta_{\pm\mathrm{KN}}$ respectively. 

\begin{figure}[htp!]
   \centering
    \includegraphics[height=0.225\textwidth]{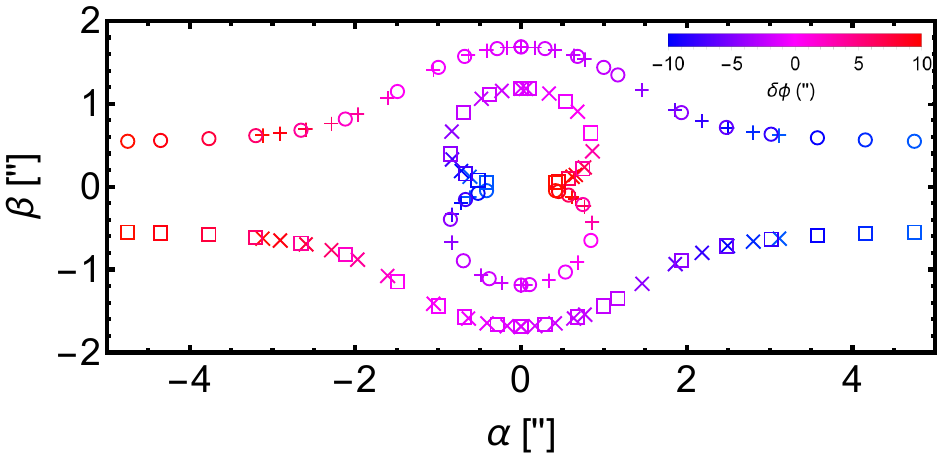}\\
   (a)\\
\includegraphics[width=0.225\textwidth]{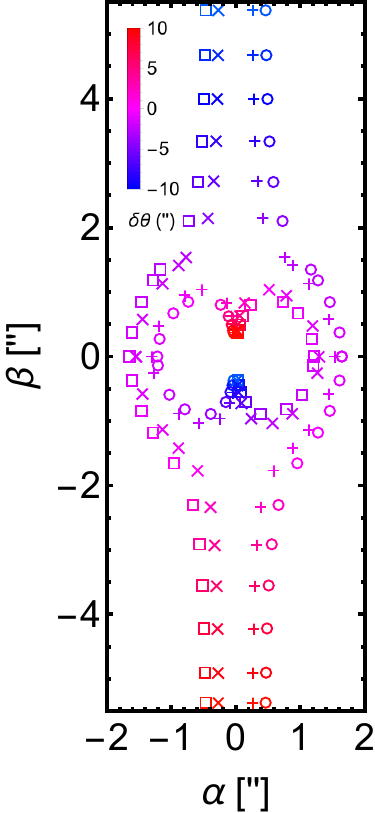}\\
   (b)\\
\includegraphics[width=0.45\textwidth]{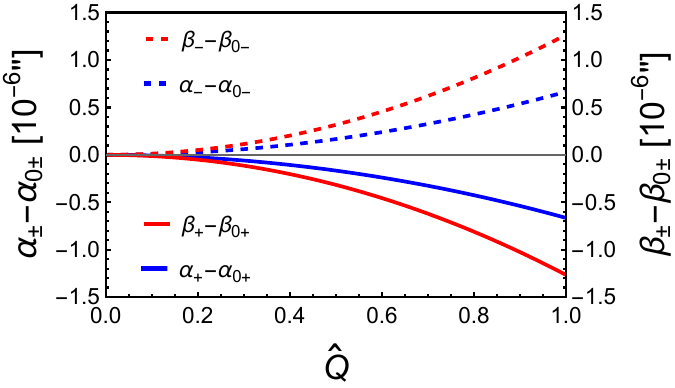}\\
   (c)
   \caption{Apparent angles of lensed images in the celestial sky in KN spacetime. (a) $\delta\phi$ varies from $-10^{\prime\prime}$ to $10^{\prime\prime}$ with fixed $\delta\theta=1^{\prime\prime}$ ($\Box$ and $\times$) and $-1^{\prime\prime}$ ($\circ$ and $+$). (b) $\delta\theta$ varies from $-10^{\prime\prime}$ to $10^{\prime\prime}$ with fixed $\delta\phi=1^{\prime\prime}$ ($\Box$ and $\times$) and $-1^{\prime\prime}$ ($\circ$ and $+$). The symbols $\Box$ and  $\circ$ are for $\theta_s=\pi/3$ and the $\times$ and $+$ are for $\theta_s=\pi/6$ respectively. The color of the symbols from blue to red indicates that the changing angles increase from $-10^{\prime\prime}$ to $10^{\prime\prime}$.  (c) Variation of the apparent angles as $\hatcq$ increases. In all plots, $\hata=\hatcq=1/2,\,v=1,\,M=4.1\times 10^6M_\odot,\,r_s=r_d=8.34$ kpc are used. In (c), $\delta \theta=10^{-4\prime\prime},\,\delta \phi=10^{-4\prime\prime},\,\theta_s=\pi/6$. The $\alpha_{0+}=0.60780640^{\prime\prime},\,\alpha_{0-}=-0.60783265^{\prime\prime},\,\beta_{0+}=1.2776103^{\prime\prime},\beta_{0-}=-1.2776603^{\prime\prime}$.
   }
   \label{fig:KNQab}
\end{figure}

In Fig. \ref{fig:KNQab} the angular positions of the GL images as functions of $\delta\phi$ (a) and $\delta\theta$ (b) are plotted. It is seen that as $\delta\phi$ varies from $0$ to $10^{\prime\prime}$ while keeping $\delta\theta$ at $1^{\prime\prime}$ the two images are in the first and third quadrants respectively. The image in the third quadrant is separated further from the lens than the one in the first quadrant. Since in this parameter settings, the effect of spin $\hata$ on the apparent angles of the images is weak, when we flip $\delta\phi$ to the range of $-10^{\prime\prime}$ to 0, or $\delta\theta$ to $-1^{\prime\prime}$, the images are reflected by the $y$ and $x$ axes respectively in the 2d celestial frame. What is more interesting is the effect of $\theta_s$ in this plot. It is seen that the trace of the images as $\delta\phi$ varies for $\theta_s=\pi/6$ almost coincides with that for $\theta_s=\pi/3$. The reason can be understood from the fact that when the spin $\hata$ effect is not strong, the total deflection in an SAS spacetime is approximately the same as in an SSS spacetime. In SSS spacetimes, $\theta_s$ only characterize the altitude of the images while when $\delta\phi$ scans through a range, the traces of the images will coincide if the origin of the local 2d celestial sky is allowed to shift, as in the case of Fig. \ref{fig:KNQab} (a). 
For the variation of $\delta\theta$ with fixed $\delta\phi=\pm 1^{\prime\prime}$, different $\theta_s$ however allows a different contribution from $\delta\theta$ to the total deflection $\delta\eta$, which roughly equals
\begin{align}
\delta\eta\approx \sqrt{\delta\theta^2+\sin^2\theta_s\delta\phi^2}.
\end{align}
And therefore the image traces will not coincide, as shown in Fig. \ref{fig:KNQab} (b). 

Note that in both plots, we have set $\hata=1/2,\,\hatcq=1/2$. The effect of these parameters under the current parameter settings, as seen from Eq. \eqref{eq:aaKNapp}, is very small compared to the apparent angles themselves, and therefore can not be recognized in plots (a) and (b). Therefore in (c), we show the small variation of the apparent angles $(\alpha_{\pm \mathrm{KN}},\,\beta_{\pm \mathrm{KN}})$ as $\hatcq$ increases, where $\alpha_{0\pm}=\alpha_{\pm\mathrm{KN}}(\hatcq=0)$, $\beta_{0\pm}=\beta_{\pm \mathrm{KN}}(\hatcq=0)$. As can be seen, the sizes of both images' apparent angles decrease by about $10^{-6\prime\prime}$ as $\hatcq$ increases. Both the trend and changed amount agree with the prediction of Eq. \eqref{eq:aaKNapp}.

\subsection{Deflections and GL in  KS spacetime}

KS BH is a kind of rotating and charged black hole in the four-dimensional heterotic string theory \cite{Sen:1992ua}. Although both the strong \cite{Gyulchev:2009dx} and weak \cite{Gyulchev:2006zg} deflection limits of GL effects have been studied in this spacetime using approaches different from ours, these studies are either in the (quasi-)equatorial plane or did not express the deflections in terms of the original source and kinetic variables. Here we will extend them to the general non-equatorial case, with finite distance effect and timelike effect taken into account. 

The metric of the KS spacetime is given by \cite{Jiang:2019vww}
\begin{align}
    \mathrm{d}s^2=&-\frac{\Sigma_{\mathrm{KS}}-2Mr}{\Sigma_{\mathrm{KS}}}\mathrm{d}t^2-\frac{4Mra\sin^2\theta}{\Sigma_{\mathrm{KS}}} \mathrm{d}t\mathrm{d}\phi\nn\\&\frac{\left[\left(r^2+2b r+a^2\right)^2-\Delta_{\mathrm{KS}} a^2\sin^2\theta\right]\sin^2\theta}{\Sigma_{\mathrm{KS}}} \mathrm{d}\phi^2\nn\\&+\frac{\Sigma_{\mathrm{KS}}}{\Delta_{\mathrm{KS}}}\mathrm{d}r^2+\Sigma_{\mathrm{KS}} \mathrm{d}\theta^2,\label{eq:ksmetric}
\end{align}
where 
\begin{align}
&\Sigma_{\mathrm{KS}}=r(r+2b)+a^2\cos^2\theta,\nn\\&\Delta_{\mathrm{KS}}=r(r+2b)-2Mr+a^2,\nn
\end{align}   
with $b\equiv Q^2/(2M)\geq0$. This metric reduces to the Kerr one when $b=0$. The metric functions satisfy the separation conditions \eqref{eq:sepcond} too and the corresponding functions and their asymptotic expansions are
\begin{subequations}
\label{eq:ksmexp}
\begin{align}
    A^{(r)}_{\mathrm{KS}}=&-\frac{a^2}{4\Delta_{\mathrm{KS}}}=-\frac{a^2}{4r^2}-\frac{a^2(M-b)}{2r^3}+\mathcal{O}(r)^{-4},\\
    B^{(r)}_{\mathrm{KS}}=&-\frac{M a r}{\Delta_{\mathrm{KS}}}=-\frac{aM}{r}-\frac{2aM(M-b)}{r^2}+\mathcal{O}(r)^{-3},\\
    C^{(r)}_{\mathrm{KS}}=&\frac{(r^2+2b r+a^2)^2}{4\Delta_{\mathrm{KS}}}\nn\\=&\frac{r^2}{4}+\frac{(M+b)r}{2}+\frac{a^2+4M^2}{4}+\mathcal{O}(r)^{-1},\\
    \mathcal{D}(r)_{\mathrm{KS}}=&r^2-2(M-b)r+a^2,\\
    G^{(r)}_{\mathrm{KS}}=&r^2+2br.
\end{align}
\end{subequations}

Substituting the coefficients Eqs. \eqref{eq:ksmexp} into Eqs. \eqref{eq:delta phi2} and \eqref{eq:delta theta2} and carrying out the small $p_{s,d}$ expansion, the deflection angles in KS spacetime are found as
\begin{widetext}
\begin{subequations}
\label{eq:dks}
\begin{align}
    \Delta\phi_{\mathrm{KS}}=&s_2\pi+\frac{4\hat{a}M^2}{vr_0^2}-\frac{8s_{m}^2\hat{a}M^2}{s_s^2vr_0^2}+\frac{s_2s_{m}}{s_s^2}\left[\frac{2(1+v^2)M}{v^2r_0}-\left(p_s+p_d\right)-\frac{\zeta_{\mathrm{KS}}M^2}{4v^4r_0^2}-\frac{M\left(p_s+p_d\right)}{v^2r_0}\right]\nn\\&+\frac{s_1s_2s_{m}c_s\sqrt{c_m^2-c_s^2}}{s_s^4}\left[\left(p_s+p_d\right)-\frac{2(1+v^2)M}{v^2r_0}\right]^2+\mathcal{O} (\epsilon)^3,\label{eq:delta phi KS}\\
    \Delta\theta_{\mathrm{KS}}=&\frac{s_1\sqrt{c_m^2-c_s^2}}{s_s}\Bigg[\frac{2(1+v^2)M}{v^2r_0}-\left(p_s+p_d\right)-\frac{\left(\zeta_{\mathrm{KS}}+32s_2s_{m}v^3\hat{a}\right)M^2}{4v^4r_0^2}-\frac{M\left(p_s+p_d\right)}{v^2r_0}\Bigg]\nn\\&-\frac{c_ss_{m}^2}{2s_s^3}\left[\left(p_s+p_d\right)-\frac{2(1+v^2)M}{v^2r_0}\right]^2+\mathcal{O} \left(\epsilon\right)^3,\label{eq:delta theta KS}
\end{align}
\end{subequations}
\end{widetext}
where
\begin{align}
    \zeta_{\mathrm{KS}}=&8+8v^2-12\pi v^2-3\pi v^4+\frac{4v^4\hat{b}\left(p_s+p_d\right)r_0}{M}\nn\\&+(8v^2+4\pi v^2+8v^4+2\pi v^4)\hat{b}+\pi v^4 \hat{b}^2,\label{zetaKS}
\end{align}
and $\hat{b}\equiv b/M$.
Comparing Eq. \eqref{eq:dks} to the corresponding results in Eq. \eqref{eq:dkn} in KN spacetime, we see that the only change is essentially in the definition of $\zeta_\mathrm{KS}$. 
This is understandable because both the KN and KS spacetimes reduce to the Kerr one if we set $Q=0$ in the former and $b$ (or $Q$) in the latter, and these two parameters only appear from the second order in each of the deflections $\Delta\phi$ and $\Delta\theta$. 
The infinite source/detector limit and null limit of the deflections \eqref{eq:dks} can be easily obtained by putting $v=1$ and $p_{s,d}=0$ respectively. The equatorial plane case of $\Delta\phi$ was found in Eq. (49) of Ref. \cite{Gyulchev:2009dx} for light ray and infinite source/detector and agrees with our result under these limits.

\begin{figure}[htp!]
   \centering
\includegraphics[width=0.45\textwidth]{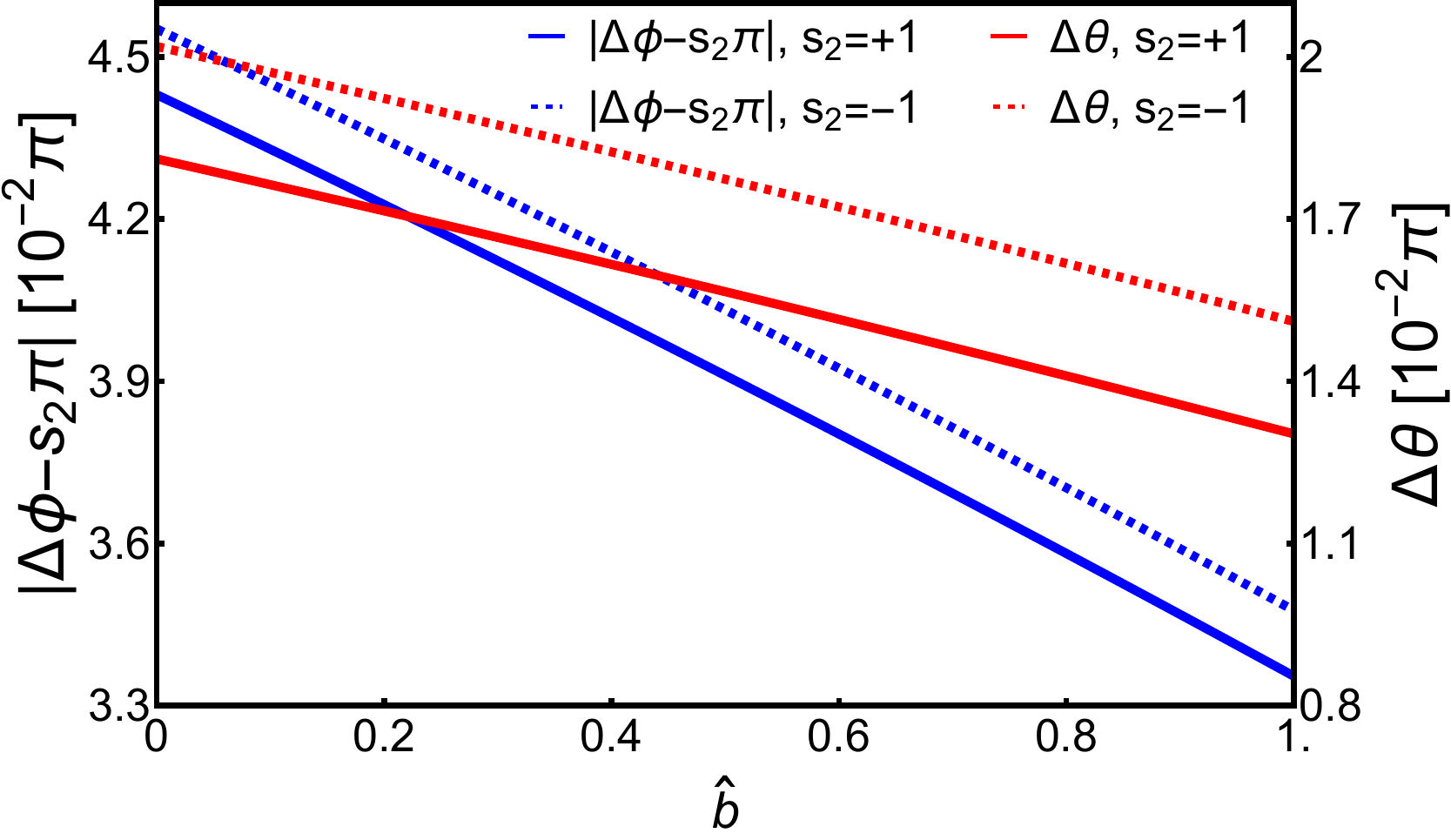}\\
(a)\\
\includegraphics[width=0.45\textwidth]{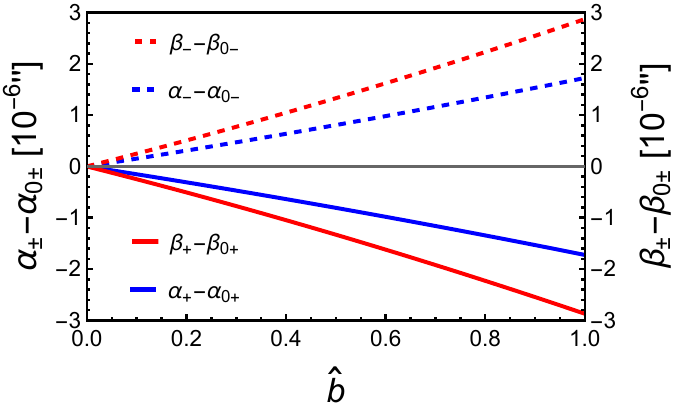}\\
(b)
   \caption{The dependences of $\Delta\phi$ and $\Delta\theta$ (a) and $\alpha$ and $\beta$ on $\hatb$ in KS spacetime. In (a),   $v=1,\,\hata=1/2,\,r_s=r_d=400M,\,\theta_s=\pi/4,\,\theta_m=\pi/5,\,r_0=20M$ were used. In (b), $v=1,\,\hata=1/2,\,M=4.1\times 10^6M_\odot,\,r_s=r_d=8.34~\mathrm{kpc},\,\delta \theta=10^{-4\prime\prime},\,\delta \phi=10^{-4\prime\prime},\,\theta_s=\pi/6$ were used.
}
\label{fig:ksdef}
\end{figure}

To study the effect of the new parameter $\hatb$ on the deflections, in Fig. \ref{fig:ksdef} (a) we plot the dependence of $\Delta\phi_\mathrm{KS}$ and $\Delta\theta_\mathrm{KS}$ on $\hatb$. It is seen that as $\hatb$ increases, the both deflections $\Delta\phi_\mathrm{KS}$ and $\Delta\theta_\mathrm{KS}$ for all spin $\hata$ decrease monotonically. This effect is qualitatively similar to the effect of $\hatcq^2$ in KN spacetime. 

With the deflection known, we can solve the GL equations \eqref{eq:gleq} for $r_0$ and $\theta_m$ too. Again, if we use the deflection angles of high enough order such that the effect of $\hatb$ is taken into account, then these GL equations are polynomials whose solutions are too lengthy to present here and therefore we will not do so. We also studied the dependence of the solved $(r_0,\,\theta_m)$ on $\hatb$ and found that it is also qualitatively similar to the effect of $\hatcq^2$ in KN spacetime. That is, the $r_0$ is decreased as $\hatb$ increases for fixed $\hata$ while $|\cos\theta_m|$ is increased but only very weakly in the WDL. 

Using these $r_0$ and $\theta_m$ in Eq. \eqref{eq:apparentangles}, the apparent angles in the KS spacetime are found as
\begin{subequations}
\begin{align}
    \alpha_{\pm\mathrm{KS}}=&\sin^{-1}\frac{L_\pm\left(\Delta_{\mathrm{KS}}-a^2\sin^2\theta\right)+2aMEr\sin^2\theta}{\sin{\theta}\sqrt{\Delta_{\mathrm{KS}}\Sigma_{\mathrm{KS}}\left[\Sigma_{\mathrm{KS}} E^2+\kappa\left(\Sigma_{\mathrm{KS}}-2M r\right)\right]}}\Bigg|_d,\\
    \beta_{\pm \mathrm{KS}}=&s_1\sin^{-1}\sqrt{\frac{\Theta_\pm\left(\Sigma_{\mathrm{KS}}-2M r\right)}{\Sigma_{\mathrm{KS}}\left[\Sigma_{\mathrm{KS}} E^2+\kappa\left(\Sigma_{\mathrm{KS}}-2M r\right)\right]}}\Bigg|_d,
\end{align}
\end{subequations}
where the $\Theta_\pm$ in the KS spacetime also takes the form of Eq. \eqref{eq:thetakn} but its $(L_\pm, \,K_\pm)$ have different relations to $(r_{0\pm},\,\theta_{m\pm})$ through Eqs. \eqref{eq:solved L} and \eqref{eq:solved GCC}. For null rays and the small $M/r_0$ and $M/r_d$ limit, these apparent angles are approximated as
\begin{subequations}
\label{eq:aaksapp}
\begin{align}
    \alpha_{\pm\mathrm{KS}}\simeq&\frac{s_2s_{m\pm}}{s_d\hat{r}_d}\left(\hat{r}_{0\pm}+\hat{b}+1+\frac{3+\hat{a}^2- \hat{b}^2-2\hat{b}-4s_2s_{m\pm}\hat{a}}{2\hat{r}_{0\pm}}\right),\label{eq:alphaKSappro}\\
    \beta_{\pm \mathrm{KS}}\simeq&\frac{s_1\sqrt{s_d^2-s_{m\pm}^2}}{s_d\hat{r}_d}\nn\\&\times\left(\hat{r}_{0\pm}+\hat{b}+1+\frac{3+\hat{a}^2 c_d^2- \hat{b}^2-2\hat{b}-4s_2s_{m\pm}\hat{a}}{2\hat{r}_{0\pm}}\right).\label{eq:betaKSappro}
\end{align}
\end{subequations}
In contrast to the parameter $\hatcq$ in the KN apparent angles \eqref{eq:aaKNapp}, we see that the parameter $\hatb$ affects the apparent angles in KS spacetime from the order $b/r_d$ explicitly. However, we also point out that as $\hatb$ increases from zero, its influence on $r_{0\pm}$ is also at this order but slightly larger and with an opposite sign. 

Fig. \ref{fig:ksdef} (b) shows the apparent locations of the images in the KS spacetime as $\hatb$ varies. The $\alpha_{0\pm}$ and $\beta_{0\pm}$ values are the same as in Fig. \ref{fig:KNQab} (c) since at $\hatb=0$, the KS spacetime reduces to the Kerr one. It is seen that as argued above, the total effect of $\hatb$ from 0 to 1 is also the decrease of the image apparent angles by about $2\times 10^{-6\prime\prime}$. In this sense, the effect of $b$ in the KS spacetime is similar to that of parameter $Q^2$ in the KN spacetime, both qualitatively and quantitatively.

\subsection{Deflections and GL in  RSV spacetime}

RSV spacetime is another modification from the Kerr spacetime that satisfies the separation conditions \eqref{eq:sepcond}. 
Its metric is given by \cite{Mazza:2021rgq}
\begin{align}
    \mathrm{d}s^2=&-\frac{\Sigma_\mathrm{R}-2M\sqrt{r^2+l^2}}{\Sigma_\mathrm{R}}\mathrm{d}t^2-\frac{4Ma\sin^2\theta\sqrt{r^2+l^2}}{\Sigma_\mathrm{R}}\mathrm{d}t\mathrm{d}\phi\nn\\&+\frac{\left[(r^2+l^2+a^2)^2-\Delta_\mathrm{R} a^2\sin^2\theta\right]\sin^2\theta}{\Sigma_\mathrm{R}}\mathrm{d}\phi^2\nn\\&+\frac{\Sigma_\mathrm{R}}{\Delta_\mathrm{R}}\mathrm{d}r^2+\Sigma_\mathrm{R} \mathrm{d}\theta^2,\label{eq:rsvmetric}
\end{align}
where
\begin{align}
&\Sigma_\mathrm{R}=r^2+l^2+a^2\cos^2\theta,\nn\\&\Delta_\mathrm{R}=r^2+l^2+a^2-2M\sqrt{r^2+l^2},\nn
\end{align}
The parameter $l\geq0$ is a length scale responsible for the regularization of the central singularity.
When $l=0$, this reduces to the Kerr spacetime too.  It can also describe two-way traversable
wormhole $(l>2M)$, a one-way wormhole $(l=2M)$, and regular black hole $(l<2M)$ at different values of $l$ \cite{Mazza:2021rgq}. The GL effects of null rays with source/detector at infinite distance in the equatorial plane were studied in the strong
deflection limit in this spacetime in Ref. \cite{Islam:2021ful}. The WDL deflection angle of the null signal on the equatorial plane without the finite distance effect was obtained using the Gauss-Bonnet theorem method in Ref. \cite{Gao:2023ipv}. 

\begin{figure}[htp!]
   \centering
\includegraphics[width=0.45\textwidth]{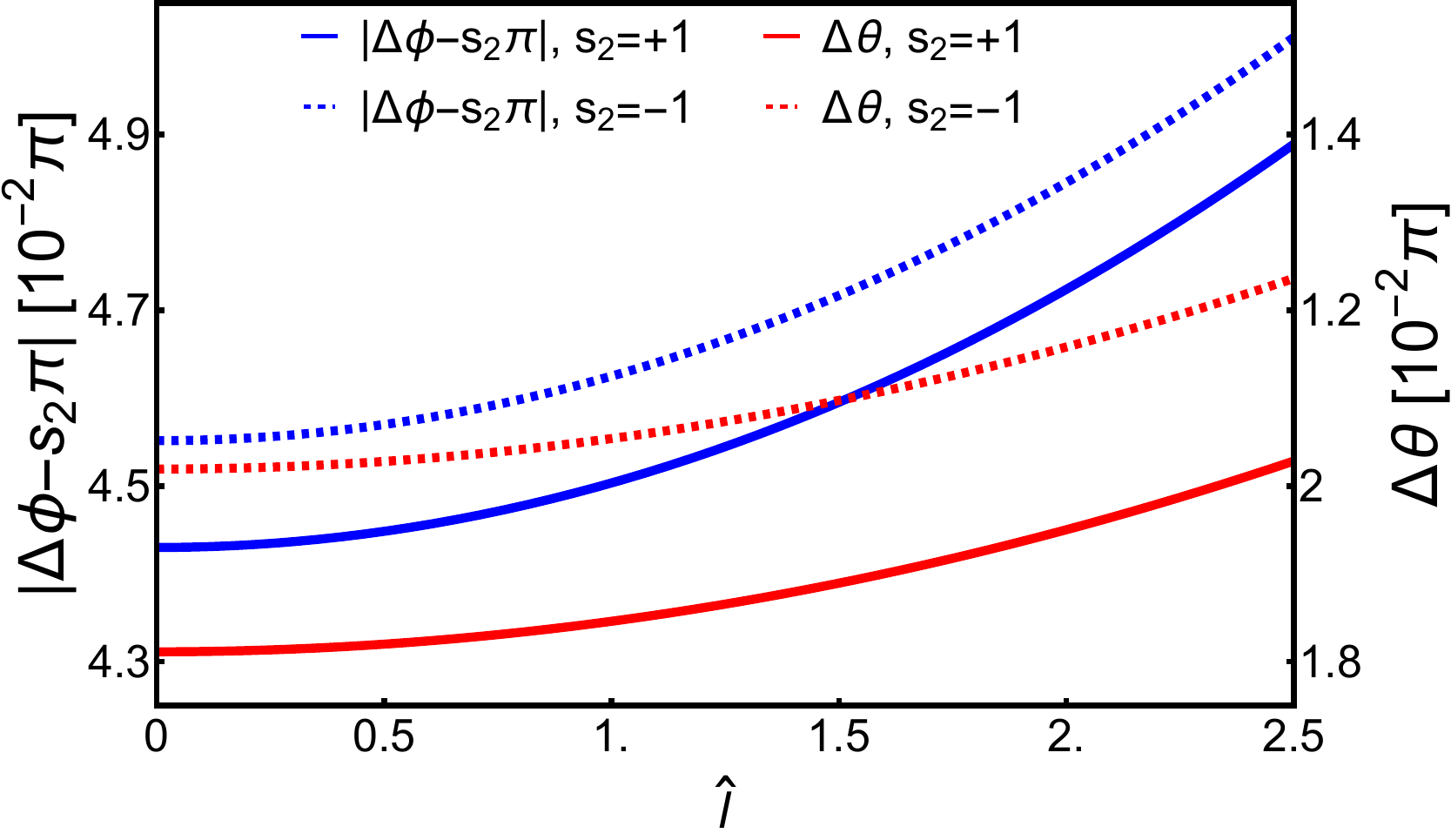}\\
(a)\\
\includegraphics[width=0.45\textwidth]{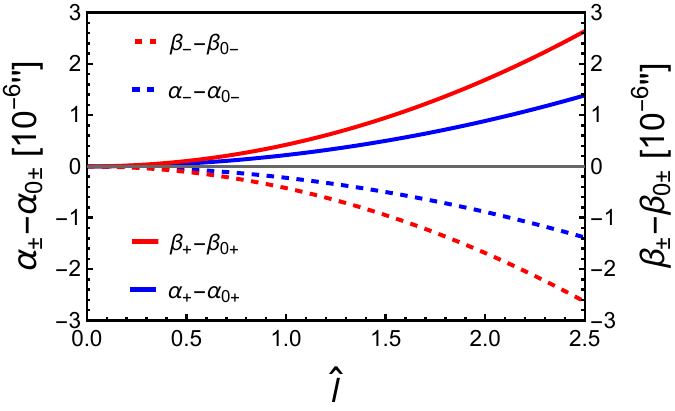}\\
(b)
\caption{The dependences of $\Delta\phi$ and $\Delta\theta$ (a) 
and $\alpha$ and $\beta$ (b) on $\hatl$ in RSV spacetime. In (a),  $r_s=r_d=400M,\,\theta_m=\pi/5,\,r_0=20M,\,\theta_s=\pi/4$ are used to see clearly the effect. In (b),  $M=4.1\times 10^6M_\odot,\,r_s=r_d=8.34~\mathrm{kpc},\, \delta \theta=10^{-4\prime\prime},\,\delta \phi=10^{-4\prime\prime},\,\theta_s=\pi/6$. In both plots, $v=1,\,\hata=1/2$ are used.
   \label{fig:rsvdef}}
\end{figure}

The asymptotic expansions of separated functions associated with the metric are
\begin{subequations}
    \begin{align}
     A^{(r)}_\mathrm{R}=&-\frac{a^2}{4\Delta_\mathrm{R}}=-\frac{a^2}{4 r^2}-\frac{Ma^2}{2 r^3}+\mathcal{O}(r)^{-4},\\
    B^{(r)}_\mathrm{R}=&-\frac{M a \sqrt{r^2+l^2}}{\Delta_\mathrm{R}}=-\frac{Ma}{r}-\frac{2M^2a}{r^2}+\mathcal{O}(r)^{-3},\\
    C^{(r)}_\mathrm{R}=&\frac{(r^2+a^2+l^2)^2}{4\Delta_\mathrm{R}}\nn\\=&\frac{r^2}{4}+\frac{Mr}{2}+\frac{a^2+4M^2+l^2}{4}+\mathcal{O}(r)^{-1},\\
    \mathcal{D}(r)_\mathrm{R}=&r^2-2Mr+a^2+l^2+\mathcal{O}(r)^{-1},\\
    G^{(r)}_\mathrm{R}=&r^2+l^2.
    \end{align}
\end{subequations}
Substituting the coefficients in these functions into Eqs. 
\eqref{eq:delta phi2} and \eqref{eq:delta theta2} and after the small $p_{s,d}$ expansion, the deflection angles in the RSV spacetime becomes
\begin{widetext}
\begin{subequations}
\label{eq:drsv}
\begin{align}
    \Delta\phi_{\mathrm{R}}=&s_2\pi+\frac{4\hat{a}M^2}{vr_0^2}-\frac{8s_{m}^2\hat{a}M^2}{s_s^2vr_0^2}+\frac{s_2s_{m}}{s_s^2}\left[\frac{2(1+v^2)M}{v^2r_0}-\left(p_s+p_d\right)-\frac{\zeta_{\mathrm{R}}M^2}{4v^4r_0^2}-\frac{M\left(p_s+p_d\right)}{v^2r_0}\right]\nn\\&+\frac{s_1s_2s_{m}c_s\sqrt{c_m^2-c_s^2}}{s_s^4}\left[\left(p_s+p_d\right)-\frac{2(1+v^2)M}{v^2r_0}\right]^2+\mathcal{O} (\epsilon)^3,\label{eq:delta phi R}\\
    \Delta\theta_{\mathrm{R}}=&\frac{s_1\sqrt{c_m^2-c_s^2}}{s_s}\Bigg[\frac{2(1+v^2)M}{v^2r_0}-\left(p_s+p_d\right)-\frac{\left(\zeta_{\mathrm{R}}+32s_2s_{m}v^3\hat{a}\right)M^2}{4v^4r_0^2}-\frac{M\left(p_s+p_d\right)}{v^2r_0}\Bigg]\nn\\&-\frac{c_ss_{m}^2}{2s_s^3}\left[\left(p_s+p_d\right)-\frac{2(1+v^2)M}{v^2r_0}\right]^2+\mathcal{O} \left(\epsilon\right)^3,\label{eq:delta theta R}
\end{align}
\end{subequations}
\end{widetext}
where
\begin{align}
    \zeta_\mathrm{R}=8+8v^2-12\pi v^2-3\pi v^4-\pi v^4\hat{l}^2,\label{zetaRSV}
\end{align}
and $\hat{l}\equiv l/M$. The reduction of Eq. \eqref{eq:delta phi R} on the equatorial plane for null rays with infinite source/detector distance agrees with Eq. (21) of \cite{Gao:2023ipv}. 
Similar to KS deflections \eqref{eq:dks}, the deflections \eqref{eq:drsv} are different from the KN case result \eqref{eq:dkn} also in its $\zeta_R$. However unlike the $\hatcq^2$ in $\zeta_\mathrm{KN}$ and $\hatb$ in $\zeta_\mathrm{KS}$, here the regularization length scale $\hatl^2$ has a negative sign in 
$\zeta_\mathrm{R}$ and therefore its effects on the deflections $(\Delta \phi_\mathrm{R},\, \Delta \theta_\mathrm{R})$, the $(r_0,\,\theta_m)$ and the apparent angles $(\alpha_{\pm\mathrm{R}},\,\beta_{\pm\mathrm{R}})$ in Eq. \eqref{eq:aarsvapp} are all opposite to those two parameters, as we will show in Fig. \ref{fig:rsvdef} (a) and (b) respectively.  

After solving $(r_0,\,\theta_m)$ and substituting into Eq.  \eqref{eq:apparentangles}, the 
apparent angles in the RSV spacetime are found to be
\begin{subequations}
    \begin{align}
    &\alpha_{\pm\mathrm{R}}\nn\\
    =&\sin^{-1}\frac{L_\pm\left(\Delta_\mathrm{R}-a^2\sin^2\theta\right)+2aME\sqrt{r^2+l^2}\sin^2\theta}{\sin{\theta}\sqrt{\Delta_\mathrm{R}\Sigma_\mathrm{R}\left[\Sigma_\mathrm{R} E^2+\kappa\left(\Sigma_\mathrm{R}-2M\sqrt{r^2+l^2}\right)\right]}}\Bigg|_d,\\
    &\beta_{\pm\mathrm{R}}\nn\\
    =&s_1\sin^{-1}\sqrt{\frac{\Theta_\pm\left(\Sigma_\mathrm{R}-2M\sqrt{r^2+l^2}\right)}{\Sigma_\mathrm{R}\left[\Sigma_\mathrm{R} E^2+\kappa\left(\Sigma_\mathrm{R}-2M \sqrt{r^2+l^2}\right)\right]}}\Bigg|_d,
\end{align}
\end{subequations}
where $\Theta_\pm$ is still given by Eq. \eqref{eq:thetakn} while its  $(L_\pm,\, K_\pm)$ depends on $(r_{0\pm},\,\theta_{m\pm})$ through Eqs. \eqref{eq:solved L} and \eqref{eq:solved GCC}. The approximations of these apparent angles are 
\begin{subequations}
\label{eq:aarsvapp}
    \begin{align}
    \alpha_{\pm\mathrm{R}}\simeq&\frac{s_2s_{m\pm}}{s_d\hat{r}_d}\left(\hat{r}_{0\pm}+1+\frac{3+\hat{a}^2+\hat{l}^2-4s_2s_{m\pm}\hat{a}}{2\hat{r}_{0\pm}}\right),\label{eq:alphaRSVappro}\\
    \beta_{\pm\mathrm{R}}\simeq&\frac{s_1\sqrt{s_d^2-s_{m\pm}^2}}{s_d\hat{r}_d}\left(\hat{r}_{0\pm}+1+\frac{3+\hat{a}^2 c_d^2+\hat{l}^2-4s_2s_{m\pm}\hat{a}}{2\hat{r}_{0\pm}}\right).\label{eq:betaRSVappro}
\end{align}
\end{subequations}

In Fig. \ref{fig:rsvdef} (b), we plotted the dependences of the deflections $(\Delta \phi_\mathrm{R},\, \Delta \theta_\mathrm{R})$ and the apparent angles $(\alpha_{\pm\mathrm{R}},\,\beta_{\pm\mathrm{R}})$ on $\hatl$. Unlike the effect of $\hatcq$ in KN spacetime, here $\hatl$ increases $r_{0\pm}$ and consequently the apparent angles of the images. The amount of reduction of either $\alpha_{\pm R}$ or $\beta_{\pm R}$ as $\hatl$ grows to 2.5 is comparable to that for $\hatcq$ in the KN case or $\hatb$ in the KS case.

\section{Conclusions and Discussions\label{sec:sumdis}}

In this work, we studied the off-equatorial deflections and GL of both null and timelike signals in general SAS spacetimes in the weak deflection limit, with the finite distance effect of the source and detector. It is found that as long as the metric functions satisfy certain common separable variable conditions \eqref{eq:sepcond}, which allows the existence of a GCC, the deflection angles in both the $\phi$ and $\theta$ directions can always be found using the perturbative method. The results, as shown in Eqs. \eqref{eq:delta phi2} and \eqref{eq:delta theta2}, are dual series of $M/r_0$ and $r_0/r_{s,d}$, and can be directly used in a set of exact GL equations \eqref{eq:gleq}. These equations are then solved to find the apparent angles of images in such spacetimes \eqref{eq:apparentangles}. 

These results are then applied to the KN, KS, and RSV spacetimes to validate the correctness of the method and results, and to find the effect of the spacetime spin as well as that of the characteristic parameter (typically an effective charge) of these spacetimes. It is found that generally both the spacetime spin and charge appear in the second order of both $\Delta\phi$ and $\Delta\theta$, while the non-equatorial effect shows up from the very leading nontrivial order, as illustrated in Eqs. \eqref{eq:dkn}, \eqref{eq:dks} and \eqref{eq:drsv}. 

For the image apparent angles, again both the spacetime spin and (effective) charge appear in the subleading, as manifested in Eqs. \eqref{eq:aaKNapp}, \eqref{eq:aaksapp} and \eqref{eq:aarsvapp}. Therefore these parameters are quite hard to detect from the apparent angles in relativistic GL in the WDL. 

To show the generality of our method, we supplemented a few other spacetimes whose off-equatorial deflections can be found using our method in Appendix \ref{sec:appendix2}.
We summarize the results computed in the main text and this appendix in Tab. \ref{tab:one} to clearly present the results and the effect of the main parameter(s) in the spacetime on the deflection and/or apparent angles.

\begin{widetext}
\begin{table}[htp!]
\centering
\makebox[\textwidth][c]{

\begin{tabular}{l|c|c|c|c|c|c}
\hline
\hline
Spacetime &Metric Eq. & Para. & Def. angle Eq. & Order & App. angle Eq. & Para. effect \\
\hline
\hline
Kerr-Newmann  &\eqref{eq:knmetric} & $Q$ & \eqref{eq:dkn}& $2$&\eqref{eq:aaKNapp}&$\searrow$ \\
\hline
Kerr-Sen &\eqref{eq:ksmetric} & $b$ & \eqref{eq:dks} &$2$&\eqref{eq:aaksapp}&$\searrow$ \\
\hline
Rotating Simpson-Visser &\eqref{eq:rsvmetric} & $l$ & \eqref{eq:drsv} &$2$&\eqref{eq:aarsvapp}&$\nearrow$ \\
\hline
Rotating Bardeen &\eqref{eq:leclass} with \eqref{eq:mbard}& $g$ & \eqref{eq:defapp} with \eqref{eq:mbard}&3 or higher&\multicolumn{2}{c}{} \\
\cline{1-5}
Rotating Hayward &\eqref{eq:leclass} with \eqref{eq:mhayward}& $k$ & \eqref{eq:defapp} with \eqref{eq:mhayward}&3 or higher&\multicolumn{2}{c}{} \\
\cline{1-5}
Rotating Ghosh &\eqref{eq:leclass} with \eqref{eq:mghosh}& $h$ & \eqref{eq:defapp} with \eqref{eq:mghosh}&$2$&\multicolumn{2}{c}{} \\
\cline{1-5}
Rotating Tinchev &\eqref{eq:leclass} with \eqref{eq:mtin}& $j$ & \eqref{eq:defapp} with \eqref{eq:mtin}&3 or higher&\multicolumn{2}{c}{} \\
\cline{1-5}
\cline{1-5}
\end{tabular}

}
\begin{minipage}{\textwidth}
  \centering
\caption{ Spacetime and their off equatorial deflections. From the second to last columns are the metric equation number, the main parameter of the spacetime, the deflection angles in that spacetime, the lowest order in the deflection angle from which the main parameter appears, and the equation number of the image apparent angles for the spacetime studied in the main text, and lastly the effect of the parameter on the apparent angles, i.e. the monotonicity of the apparent angle as the parameter increases. \label{tab:one}}
\end{minipage}
\end{table}
\end{widetext}

The results of this work can in principle be applied to any spacetime with a (non-spherical) axisymmetry. Inside the solar system however, the only known object that can bend the light is the Sun, and yet its dimensionless spin parameter $\hata$ is only of order $10^{-20}$ \cite{Fienga:2019uds}. Therefore the effect of the spin or the off-equatorial plane effect in the deflection angle and/or apparent angle of images can not be observed for the Sun in the foreseeable future. 
Instead, most spacetimes studied in this work are {\it black hole} spacetimes. Therefore the results are more applicable to more extreme rotating black holes, some particular examples include the M87* \cite{EventHorizonTelescope:2019dse} and the Sgr A* \cite{EventHorizonTelescope:2022wkp}, whose spin parameter $\hat{a}$ could potentially reach a much larger order (roughly a fraction of one). One can also assume that such black holes carry extra parameters such as those appearing in the Kerr-Sen or rotating-Simpson-Visser spacetimes, and attempt to use future observations to constrain the corresponding parameters.

There are a few potential extensions to this work that can be explored. The first is to study the magnification and time delays of the images in the off-equatorial GL. In Ref. \cite{Jiang:2023yvo} for the Kerr spacetime, it is known that the spacetime spin might have a stronger effect on the time delay than on image locations. The second is that we might also attempt to study the off-equatorial deflection of charged particles in electromagnetic fields. However there, the separation condition \eqref{eq:sepcond} has to be re-investigated.  

\acknowledgments

This work is supported by the National Natural Science Foundation of China. The authors thank the helpful discussion with Tingyuan Jiang and Xiaoge Xu. The work of X.Ying is partially supported by the Undergraduate Training Programs for Innovation and Entrepreneurship of Wuhan University.

\appendix

\section{Higher order items of series\label{sec:appendix1}}

Here we list some higher-order coefficients in the series appearing in the main text of the \ref{sec:The Perturbative Method}. For Eq. \eqref{eq:expansion2}, we present two more coefficients, which are also used in the computations in the main text
\begin{align}
    &m_{r,3}=\frac{p^2}{8\sqrt{(1-p^2)(\kappa g_0+4E^2c_0)d_0^5}}\bigg\{4d_0d_2+16a_2d_0^2s_m^2\nn\\&-\frac{8d_0d_1(\kappa g_1+4E^2c_1)}{(1+p)(\kappa g_0+4E^2c_0)}-\frac{16s_2b_2d_0^2s_mE}{(1+p)\sqrt{\kappa g_0+4E^2c_0}}\nn\\&-\frac{3\left[ d_0\left(\kappa g_1+4E^2 c_1\right)-d_1(1+p)\left(\kappa g_0+4E^2 c_0\right)\right]^2}{(1+p^2)(\kappa g_0+4E^2c_0)^2}\bigg\},\\
&m_{\theta,3}=\frac{C_2+C_3 c^2-C_1 c^4}{2(c_m^2-c^2)(\kappa g_0+4E^2c_0)^{3/2}},
\end{align}
where
\begin{align}
   C_1=&a^2(E^2+\kappa),\nn\\
   C_2=&\bigg[\kappa g_2+4E^2c_2-4a_2s_m^2(\kappa g_0+4E^2 c_0)-a^2E^2-s_m^2 C_1\nn\\&+4s_2s_mb_2E\sqrt{\kappa g_0+4E^2 c_0}-\frac{3(\kappa g_1+4E^2c_1)^2}{4(\kappa g_0+4E^2 c_0)}\bigg]c_m^2,\nn\\
   C_3=&c_m^2 C_1-\frac{C_2}{c_m^2}.\nn
\end{align}
The corresponding integral coefficients are
\begin{align}
    &M_{r,3}=-\frac{1}{16(d_0C_4)^{5/2}}\bigg\{2d_0^2C_5^2\bigg[\frac{4+5p_j}{1+p_j^2}\sqrt{1-p_j^2}-3\cos^{-1}(p_j)\bigg]\nn\\&-4d_0d_1C_4C_5\frac{(1+p_j)^2}{1+p_j^2}\bigg[\frac{2+p_j}{1+p_j}\sqrt{1-p_j^2}-\cos^{-1}(p_j)\bigg]\nn\\&+\left(16a_2s_m^2d_0^2-3d_1^2+4d_0d_2\right)C_4^2\bigg[p_j\sqrt{1-p_j^2}+\cos^{-1}(p_j)\bigg]\nn\\&-32s_2s_mEb_2d_0^2C_4^{3/2}\bigg[\frac{2+p_j}{1+p_j^2}\sqrt{1-p_j^2}-\frac{1+p_j}{1+p_j^2}\cos^{-1}(p_j)\bigg]\bigg\},
\end{align}
where
\begin{align}
    C_4=\kappa g_0+4E^2c_0,\nn\\
    C_5=\kappa g_1+4E^2c_1,\nn
\end{align}
and
\begin{align}
    M_{\theta,3}=&\frac{2C_3-3C_1c_m^2}{4(\kappa g_0+4E^2c_0)^{3/2}}\bigg[\cos^{-1}\left(\frac{c_j}{c_m}\right)+\frac{s_1c_j}{\sqrt{c_m^2-c_j^2}}\bigg]\nn\\&+\frac{s_1c_j(C_1c_j^2c_m^2+2C_2)}{4c_m^2(\kappa g_0+4E^2c_0)^{3/2}\sqrt{c_m^2-c_j^2}}.
\end{align}

The second-order coefficient in Eq. \eqref{eq:cos theta d} is
\begin{align}
    &h_2=-s_1\sqrt{\kappa g_0+4E^2c_0}\sqrt{c_m^2-h_0^2}\sum_{j=s,d}M_{r,3}-\frac{h_0h_1^2}{2(c_m^2-h_0^2)}\nn\\&+\frac{(\kappa g_1+4E^2c_1)h_1}{2(\kappa g_0+4E^2c_0)}+\frac{\sqrt{c_m^2-h_0^2}}{4(\kappa g_0+4E^2c_0)}\bigg[\bigg(\frac{h_0}{\sqrt{c_m^2-h_0^2}}\nn\\&+\frac{c_s}{\sqrt{c_m^2-c_s^2}}+\frac{1}{\sqrt{d_0}}\sum_{j=s,d}\cos^{-1}(p_j)\bigg)\left(2C_3-3C_1c_m^2\right)\nn\\&+\frac{h_0\left(2C_2+C_1h_0c_m^2\right)}{c_m^2\sqrt{c_m^2-h_0^2}}+\frac{c_s\left(2C_2+C_1c_sc_m^2\right)}{c_m^2\sqrt{c_m^2-c_s^2}}\bigg].
\end{align}

\section{Applications in other SAS spacetimes\label{sec:appendix2}}
In addition to the spacetimes that we discussed in detail in Sec. \ref{sec:Applications}, many spacetimes also meet the requirements of the separation of variables we established in Sec. \ref{sec:Equations of Motion and Deflection}.
Here we briefly mentioned the off-equatorial deflections in these spacetimes. 

The following line element describe a class of spacetimes satisfying these conditions
\begin{align}
     \mathrm{d}s^2=&-\frac{\Sigma-2m(r)r}{\Sigma}\mathrm{d}t^2-\frac{4am(r)r\sin^2\theta}{\Sigma}\mathrm{d}t\mathrm{d}\phi\nn\\&+\left(r^2+a^2+\frac{2a^2m(r)r\sin^2\theta}{\Sigma}\right)\sin^2\theta \mathrm{d}\phi^2\nn\\&+\frac{\Sigma}{\Delta}\mathrm{d}r^2+\Sigma \mathrm{d}\theta^2,
     \label{eq:leclass}
\end{align}
 where
$$\Sigma=r^2+a^2\cos^2\theta,\,\Delta=r^2-2m\left(r\right)r+a^2$$
and $a,\,m(r)$ are the spacetime spin and mass functions respectively. This line element covers the Kerr spacetime when $m(r)=M$ is a constant,  the rotating Bardeen black hole \cite{Bambi:2013ufa} when 
\begin{align}
    m_{\mathrm{B}}(r)=&M\left(\frac{r^2}{r^2+g^2}\right)^{3/2}=M-\frac{3g^2 M}{2r^2}+\mathcal{O}(r)^{-3},
    \label{eq:mbard}
    \end{align}
the rotating Hayward black hole \cite{Bambi:2013ufa} when
\begin{align}
    m_{\mathrm{H}}(r)=&M\frac{r^3}{r^3+k^3}=M-\frac{k^3M}{r^3}+\mathcal{O}(r)^{-5},\label{eq:mhayward}
\end{align}
the rotating Ghosh black hole \cite{Ghosh:2014pba} when
\begin{align}
    m_{\mathrm{G}}(r)=&M e^{-h/r}=M-\frac{h M}{r}+\mathcal{O}(r)^{-2},\label{eq:mghosh}
\end{align}
and the rotating Tinchev black hole \cite{Tinchev:2015apf} when 
\begin{align}
    m_{\mathrm{T}}(r)=&M e^{-j/r^2}=M-\frac{j M}{r^2}+\mathcal{O}(r)^{-3}. \label{eq:mtin}
\end{align}
Here we have expanded $m(r)$ into the following form 
\begin{align}
    m(r)=\sum_{n=0}^\infty \frac{m_n}{r^n}
    \label{eq:mrexp}
\end{align}
and the coefficients $m_n$ for each spacetime can be easily read off from Eqs. \eqref{eq:mbard}-\eqref{eq:mtin}.
For the rotating Bardeen and Tinchev spacetimes, their characteristic parameter appears from the second order of the expansion. While the rotating Ghosh and Hayward ones appear from the first and third order respectively. 

Using the line element \eqref{eq:leclass} and mass function \eqref{eq:mrexp}, the deflection $\Delta\theta$ and $\Delta\phi$ are found as 
\begin{widetext}
\begin{subequations}
\label{eq:defapp}
\begin{align}
\Delta\phi=&s_2\pi+\frac{4\hat{a}m_0^2}{vr_0^2}-\frac{8s_m^2\hat{a}m_0^2}{s_s^2vr_0^2}+\frac{s_2s_m}{s_s^2}\left[\frac{2(1+v^2)m_0}{v^2r_0}-\left(p_s+p_d\right)-\frac{\zeta m_0^2}{4v^4r_0^2}-\frac{m_0\left(p_s+p_d\right)}{v^2r_0}\right]\nn\\&+\frac{s_1s_2s_mc_s\sqrt{c_m^2-c_s^2}}{s_s^4}\left[\left(p_s+p_d\right)-\frac{2(1+v^2)m_0}{v^2r_0}\right]^2+\mathcal{O} (\epsilon)^3,\label{eq:delta phi RBH}\\
\Delta\theta=&\frac{s_1\sqrt{c_m^2-c_s^2}}{s_s}\Bigg[\frac{2(1+v^2)m_0}{v^2r_0}-\left(p_s+p_d\right)-\frac{\left(\zeta+32s_2s_mv^3\hat{a}\right)m_0^2}{4v^4r_0^2}-\frac{m_0\left(p_s+p_d\right)}{v^2r_0}\Bigg]\nn\\&-\frac{c_ss_m^2}{2s_s^3}\left[\left(p_s+p_d\right)-\frac{2(1+v^2)m_0}{v^2r_0}\right]^2+\mathcal{O} \left(\epsilon\right)^3,\label{eq:delta theta RBH}
\end{align}
\end{subequations}
\end{widetext}
where 
\begin{align}
    \zeta=8+8v^2-12\pi v^2-3\pi v^4-\frac{2\pi v^2(2+v^2)m_1}{m_0^2}
\end{align}
and all $m_i$ should be read off from corresponding spacetime.

We also computed the deflection in the Konoplya-Zhidenko rotating non-Kerr spacetime whose metric is given by \cite{Konoplya:2016pmh}
\begin{align}
    \mathrm{d}s^2=&-\frac{N^2-W^2\sin^2\theta}{\mathcal{K}^2}\mathrm{d}t^2-2rW\sin^2\theta \mathrm{d}t\mathrm{d}\phi\nn\\&+\mathcal{K}^2r^2\sin^2\theta \mathrm{d}\phi^2+\frac{\Sigma}{N^2 r^2}\mathrm{d}r^2+\Sigma \mathrm{d}\theta^2,
\end{align}
where
\begin{subequations}
\begin{align}
&\Sigma=r^2+a^2\cos^2\theta,\,\Delta=r^2-2M r+a^2 ,\nn\\
&N^2=\frac{(\Delta-\eta/r)}{r^2},\,W=\frac{2M a}{\Sigma}+\frac{\eta a}{r^2\Sigma},\nn\\
&\mathcal{K}^2=\frac{\left(r^2+a^2\right)^2-a^2(\Delta-\eta/r)\sin^2\theta}{r^2\Sigma},\nn
\end{align}
\end{subequations}
and $\eta$ is the deformation parameter that describes the deviation from Kerr spacetime. 
However, it is found that to order $\mathcal{O}(M/r_0)^2$, the parameter $\eta$ does not appear in either of $\Delta\phi$ or $\Delta\theta$. Therefore the deflections to order $\mathcal{O}(M/r_0)^2$ in this spacetime is also given by Eq. \eqref{eq:defapp} with $m_{n\geq 1}=0$.

For future reference, we also test the applicability of our methodology to SAS but non-asymptotically flat spacetimes, such as the KN-(anti)de Sitter spacetime described by \cite{Gibbons:1977mu}
\begin{align}
    \mathrm{d}s^2=&\frac{\Delta_\theta}{\rho^2\Xi^2}\left[a\mathrm{d}t-\left(r^2+a^2\right)\mathrm{d}\phi\right]^2-\frac{\Delta_r}{\rho^2\Xi^2}\left(\mathrm{d}t-a\sin^2\theta\mathrm{d}\phi\right)^2\nn\\
    &+\rho^2\left(\frac{\mathrm{d}r^2}{\Delta_r}+\frac{\mathrm{d}\theta^2}{\Delta_\theta}\right),
\end{align}
where
\begin{align}
    &\rho^2=r^2+a^2\cos^2\theta,\,\Delta_\theta=1+\frac{\Lambda}{3}a^2\cos^2\theta,\nn\\
    &\Delta_r=\left(r^2+a^2\right)\left(1-\frac{\Lambda r^2}{3}\right)-2M r+Q^2,\,\Xi=1+\frac{\Lambda}{3}a^2,\nn
\end{align}
and Kerr-Taub-NUT spacetime with metric \cite{Miller:1973hqu}
\begin{align}
    \mathrm{d}s^2=&-\frac{\Delta-a^2\sin^2\theta}{\Sigma}\mathrm{d}t^2+\frac{2[\Delta\chi-a(\Sigma+a\chi)\sin^2\theta]}{\Sigma}\mathrm{d}t\mathrm{d}\phi\nn\\&+\frac{(\Sigma+a\chi)^2\sin^2\theta-\chi^2\Delta}{\Sigma}\mathrm{d}\phi^2+\frac{\Sigma}{\Delta}\mathrm{d}r^2+\Sigma \mathrm{d}\theta^2,
\end{align}
where
\begin{align}
    &\Sigma=r^2+(\hat{n}+a\cos\theta)^2,\nn\\
    &\Delta=r^2-2Mr+a^2-\hat{n}^2,\nn\\
    &\chi=a\sin^2\theta-2\hat{n}\cos\theta.\nn
\end{align}
It is found that they also satisfy the separation requirements \eqref{eq:sepcond} and therefore the deflection of both null and timelike rays in the equatorial or off-equatorial plane in them can be treated using our method. 
Lastly for C-type metrics which does not have the reflective symmetry about the equatorial plane, including the KN-(A)dS C-metric \cite{Anabalon:2018qfv}
\begin{align}
    \mathrm{d}s^2=&\frac{1}{H^2}\Bigg\{-\frac{f(r)}{\Sigma}\left(\frac{\mathrm{d}t}{\alpha}-a\sin^2\theta\frac{\mathrm{d}\phi}{K}\right)^2+\frac{\Sigma}{f(r)}\mathrm{d}r^2\nn\\
    &+\frac{\Sigma r^2}{h(\theta)}\mathrm{d}\theta^2+\frac{h(\theta)\sin^2\theta}{\Sigma r^2}\left[\frac{a\mathrm{d}t}{\alpha}-\left(r^2+a^2\right)\frac{\mathrm{d}\phi}{K}\right]^2\Bigg\},
\end{align}
where
\begin{align}
    &f(r)=\left(1-A^2 r^2\right)\left(1-\frac{2m}{r}+\frac{a^2+e^2}{r^2}\right)+\frac{r^2+a^2}{l^2},\nn\\
    &h(\theta)=1+2m A\cos\theta+\left[A^2\left(a^2+e^2\right)-\frac{a^2}{l^2}\right]\cos^2\theta,\nn\\
    &\Sigma=1+\frac{a^2}{r^2}\cos^2\theta,\,H=1+A r\cos\theta,\nn
\end{align}
and its subcases with $A\neq 0$, we found that only for null but not the timelike rays, the separation requirements \eqref{eq:sepcond} can be met and therefore the deflection can be studied using our approach. 
For these metrics, however, we will not list the formulas in the  $\theta$ and $\phi$ directions until more valuable applications arise.

\end{document}